\documentclass[useAMS,usenatbib]{mn2e}

\usepackage{epsfig}
\usepackage{amsmath} 

\title[Gravitational Lensing Simulations I : Covariance Matrices and Halo Catalogues]{Gravitational Lensing Simulations I : Covariance Matrices and Halo Catalogues}
\author[Joachim Harnois-D\'{e}raps,  Sanaz Vafaei and Ludovic Van Waerbeke]{Joachim Harnois-D\'{e}raps$^{1,2}$\thanks{E-mail:jharno@cita.utoronto.ca}, Sanaz Vafaei$^{3}$\thanks{E-mail:svafaei@phas.ubc.ca} and Ludovic Van Waerbeke$^{3}$\thanks{E-mail:waerbeke@phas.ubc.ca}\\
$^{1}$Canadian Institute for Theoretical Astrophysics, University of
Toronto, M5S 3H8, Canada\\
$^{2}$Department of Physics, University of Toronto, M5S 1A7, Ontario, Canada\\
$^{3}$Department of Physics and Astronomy, University of British Columbia, Vancouver, V6T 1Z1, B.C., Canada}

\begin{document}

\date{\today}

\pagerange{\pageref{firstpage}--\pageref{lastpage}} \pubyear{2011}

\maketitle

\label{firstpage}

\begin{abstract}
Gravitational lensing surveys have now become large and precise enough that the interpretation of the lensing signal
 has to take into account an increasing number of theoretical limitations and observational biases. 
 Since the lensing signal is the strongest at small angular scales, only numerical simulations can reproduce faithfully
 the non-linear dynamics and secondary effects at play. 
 This work is the first of a series in which all gravitational lensing corrections known so far will be implemented in the same set of simulations, 
 using realistic mock catalogues and non-Gaussian statistics.  In this first paper, we present the {\small TCS} simulation suite and compute  basic statistics such as the second and third order convergence and shear correlation functions. These simple tests set the range of validity of our simulations,
 which are resolving most of the signals at the sub-arc minute level (or $\ell \sim 10^{4}$). We also compute the non-Gaussian covariance matrix 
 of several statistical estimators, including many that are used in the Canada France Hawaii Telescope Lensing Survey (CFHTLenS). 
From the same realizations, we construct halo catalogues, computing a series of  properties that are required by most galaxy population algorithms.
These simulation products are publicly available for download.
\end{abstract}

\begin{keywords}
cosmology: dark matter---weak lensing---large scale structure of Universe---methods: systematic
\end{keywords}

\section{Introduction}
\label{sec:intro}

The latest measurements of  the cosmic microwave background (CMB) \citep{2011ApJS..192...14J, 2011arXiv1101.2022P}
and of large scale galaxy surveys \citep{2000AJ....120.1579Y, 2003astro.ph..6581C, 2006A&A...452...51S}  
suggest that the Universe is mostly filled with  dark energy and dark matter. 
In that so called concordance, or standard, model of cosmology, the matter that is actually observed accounts for only five per cent of the total energy.
Improving our knowledge about this dark sector is one of the biggest challenge physicists and astrophysicists are facing, 
and it was soon recognized that an international effort, which would combine complimentary techniques
such as baryonic acoustic oscillations, type 1A supernovae, weak lensing  and cluster growth,
could lead to tight constraints on dark energy parameters \citep{2006astro.ph..9591A}. 

Whereas the latest analyses \citep{2001MNRAS.327.1297P, Eisenstein:2005su, 2006PhRvD..74l3507T, 2007MNRAS.381.1053P, 2007MNRAS.381..702B, 2011ApJS..192...18K} were able to achieve per cent level precision on most parameters,
next generation surveys, including LSST \citep{2009arXiv0912.0201L}, Euclid\footnote{\tt http://www.euclid-ec.org} \citep{2011arXiv1110.3193L}, 
SKA\footnote{\tt http://www.skatelescope.org} \citep{2008AIPC.1035..303L}, Pan-STARRS\footnote{\tt http://pan-starrs.ifa.hawaii.edu/}, VST-KiDS\footnote {\tt http://www.astro-wise.org/projects/KIDS/}, DES\footnote{\tt https://www.darkenergysurvey.org/}  are designed to reach the sub-per cent level. To achieve such performance, any systematic or secondary effect needs to be understood with at least the same level of accuracy.

In the context of global dark energy effort, weak lensing analyses are particularly appreciated for their
ability to detect dark matter structures with a minimal amount of bias. 
They are based on the measurement of the degree of deformation caused by the foreground matter structures, which act as a lenses, on background light sources.
The signal allows us to characterize the average mass profile of foreground lenses, which typically consist of groups or clusters of galaxies of different type, morphology and colour, generally centred on a dark matter halo.
Although the shape of the 2-point cosmic shear signal depends on many cosmological parameters, 
it is especially powerful at constraining a combination of the normalization of the matter power spectrum $\sigma_{8}$ and the matter density $\Omega_{m}$.
The degeneracy between $\sigma_{8}$ and $\Omega_{m}$ can then be broken with measurements of the skewness and other higher-order statistics \citep{1997A&A...322....1B}.
High precision measurements of these two parameters are relevant for dark energy when combined with complimentary techniques -- 
standard ruler with baryonic acoustic oscillations, or redshift-luminosity distance with type 1A supernovae for instance  --
however it was recently shown that weak lensing is also a standalone probe.
The signal indeed has dependencies on the redshift-distance relation, the growth factor and the non-linear clustering \citep{2002PhRvD..65f3001H,2006astro.ph..9591A,2008ARNPS..58...99H}. 

In order to match the statistical and systematic accuracy of upcoming surveys,
it is thus of the utmost importance to minimize all of the theoretical uncertainties associated with 
the weak lensing technique. The accuracy at which one can model this signal depends on a number of things.
First,  one must assess  the reliability of the modelled lenses and sources distributions,
which mainly depend on the accuracy of the underlying matter density field.
Next comes the calculation of the propagation of light, which can be done with varying degrees of precision.
Finally, the modelled signal depends on the accuracy of the galaxy  population algorithm and on our understanding of all secondary effects.

Early generations of analytical calculations were performed using linear theory  \citep[see][for a review]{2001PhR...340..291B},
and used, for instance, Zel'dovich approximation to produce late redshift lenses, which assumes a linear matter power spectrum. 
These are known to underestimate the amount of structure over a large dynamical range. 
 Indeed, it was shown that for sources at redshift $z = 1$, the projection of a non-linear density field on to a light cone impacts the lensing signal
 at angles up to a few tens of arc minutes  \citep{2000ApJ...530..547J}.
Better results were obtained from higher-order perturbation theory \citep{2002PhR...367....1B}, within the halo model \citep{2002PhR...372....1C} 
or by using the non-linear predictions of  \cite{2003MNRAS.341.1311S}. 
Unfortunately, these models fail at recovering accurate lensing signals at small angles, largely due to inaccuracies in the non-linear calculations.
Similarly, estimates based on mock catalogs made with log-normal  densities \citep{1991MNRAS.248....1C} are fast and convenient,
and were shown to yield results very similar to N-body simulations \citep{2011A&A...536A..85H} in the trans-linear regime. 
However, it was discovered at the same time that log-normal mocks tend to overestimate the covariance for scales smaller than a few arc minutes, which are critical for our current  work.
For optimal measurements, it was soon realized that one needs to rely on accurate N-body simulations for modelling the non-linear density field \citep{1991MNRAS.251..600B, 1998ApJ...493...10P, 2000ApJ...537....1W,2000ApJ...530..547J, 2009A&A...499...31H}.

For the sake of constraining the dark matter and dark energy parameters, having an accurate theoretical model of the weak lensing
signal is not enough. In addition, the description of the statistical uncertainty associated with the measurements needs to be 
as accurate as the signal itself. Current data analyses work under the assumption that a Gaussian description of the field
is accurate enough, given the other sources of uncertainty involved in the measurements. 
Although this approximation is reasonable for existing surveys, it will no longer be adequate for the next generation of surveys,
which will be much more sensitive to the non-linear scales. 

In particular, the non-linear nature of the density field on small scales tends to correlate measurements in a different way.
For instance, the Fourier modes of the density fields grow independently 
in the linear regime, but couple together in the trans-linear regime \citep{2000Natur.406..376C}, giving rise to non-Gaussian features
in the three-dimensional matter power spectrum \citep{1999MNRAS.308.1179M, 2005MNRAS.360L..82R, Neyrinck:2006xd, 2009ApJ...700..479T, 2011arXiv1106.5548N, 2011arXiv1109.5746H}. These propagate in weak lensing measurements,
as observed in simulations \citep{ 2009MNRAS.395.2065T, 2009arXiv0905.0501D, 2009ApJ...701..945S, 2011ApJ...734...76S} 
and in the data \citep{2008ApJ...686L...1L}.
These correlations are in fact decreasing the number of degrees of freedom contained in the probed volume, 
hence reduces the  power at which one can constrain cosmology \citep{2009arXiv0905.0501D, 2010PhRvD..81l3015L}.
It is therefore essential to measure accurately this effect at a resolution that matches that of modern surveys,
i.e. at the sub-arc minute level.

Two of the main limitations of existing simulation suites is that they are either not resolving small enough scales, or they are 
limited in terms of number of realizations\footnote{Generally speaking, a covariance matrix  with $N^2$ elements will have converged
if estimated with {\it much more} than $N$ simulations. Numerical convergence tests on these matrices have shown that $N^2$ realizations
yield an error of the order of ten per cent on each element \citep{2009ApJ...700..479T, 2009arXiv0905.0501D}.}.
For example, the Coyote Universe simulation suite models many cosmological volumes 
of $1300$ $h^{-1}$ Mpc per side, organized in three series for each cosmology. The `L-series' consists of 16 realizations and
covers the lower-$k$ modes only, while the `H-series' covers the quasi-linear regime.
These two series are produced with a {\small PM} code, and their combination can probe scales 
up to $k \sim 0.43 h\mbox{Mpc}^{-1}$.
In addition, a single realization is ran with a tree-{\small PM} code, and resolves the $k\sim1$ $h \mbox{Mpc}^{-1}$ scales \citep{2010ApJ...713.1322L}.
These allow for a very accurate measurement of the mean power spectrum, but
are not adequate to achieve convergence on the covariance matrix. 
The analyses carried by  \cite{2011MNRAS.416.1045K} 
was based on the {\small SUNGLASS}  pipeline \citep{2011MNRAS.414.2235K},
in which 100 simulations were produced with $512^3$ particles and a box size of $512$ $h^{-1}$Mpc. 
For some measurements 100 realizations could be enough,  but unfortunately small scales
are not resolved well enough, and the agreement with theoretical predictions deviates by more than 10 per cent for $\ell>2000$.
What we need is to push the resolution  limits by at least an order of magnitude, reaching $\ell \sim 10000$.
 Similarly, the $1000$ realizations produced by \cite{2009ApJ...701..945S} have a low resolution compared to our needs.

At this point,  running a new series is necessary, and the requirements are a)
sub arc-minute accuracy on the lensing signal, and b) large statistics, for convergence of the covariance matrix. 
The first goal of this paper is to describe a new set of  simulations, the {\small TCS} simulation suite, that fulfils both of these requirements, 
and to provide a robust description of the non-Gaussian uncertainty on the 2- and 3-point estimators that are commonly used in weak lensing analyses.
 It is constructed from the {\small CUBEP3M} N-body code, with eight times more particles than the above mentioned {\small SUNGLASS} series, in a volume more than forty times smaller, thus probing much deeper in the non-linear regime.
Our choice of N-body code is largely motivated by the fact that in the absence of  high density regions, {\small P3M} Poisson solvers 
are much faster than tree-{\small PM} codes, and the particle-particle interactions at the sub-grid level
enhance the resolution significantly compared to {\small PM} codes.

With the forecasted accuracy of the next generation of surveys, many secondary effects, that were previously overlooked or neglected,
now need to be carefully examined, since they are likely to contribute to a large portion of the theoretical uncertainty.
We gather here the principal secondary signals.  
\begin{itemize}
\item {The impact of intrinsic alignment needs to be quantified in order to calibrate the lensing signal \citep{2004MNRAS.347..895H}. 
This effect is caused by the fact that galaxies that live in a same cluster are subject to a coherent tidal force, which tends to compress them along the direction to the centre of mass of the system \citep{1988MNRAS.232..339H, 2010MNRAS.402.2127S}.  }
\item{Another secondary effect that needs to be examined is the so-called `intrinsic alignment-lensing correlation' (sometimes referred to as `shear-ellipticity correlation' or contamination), a correlation that exists between the  intrinsic alignment of the foreground galaxies
and the shear signal of the same galaxies on background sources.
For this effect to occur, the foreground system needs to be relaxed enough such that the orientation of the foreground galaxy correlates with the tidal field it is subjected to \citep{2004PhRvD..70f3526H}.}
\item{Source clustering is another important secondary effect, 
which is caused by the fact that sources are not uniformly distributed : regions of the sky with more
sources are likely to provide a stronger weak lensing signal \citep{1998A&A...338..375B}. }
\item{Also to be tested is the possible intrinsic alignment of galaxies with voids, an effect which was previously found to be consistent with zero \citep{2006MNRAS.371..750H}.}
\end{itemize}

Some of these effects have already been studied in simulations \citep{2006MNRAS.371..750H, 2008MNRAS.388..991S,2011MNRAS.410..143S},
but the statistical accuracy and the resolution were limited, such that these previous works need to be extended.
More importantly, the secondary effects have been studied separately so far, and we do not yet understand how they blend together.
This is the second goal of this paper: we set the stage to start quantifying in details how these weak lensing secondary effects
interact, with sub-arc minute accuracy.
Because the only way to quantify their impact in the data is by measuring their combined contribution in mock galaxy catalogues, 
our long term plan is to construct a large sample of mock galaxy catalogues to quantify both the mean and 
the uncertainty on these secondary effects. 
For this, we need to test separately the accuracy of the underlying density fields, the halo finder algorithm, the galaxy population scheme and the proposed weak lensing estimator (see \cite{2007MNRAS.379.1507F} for example). 
This paper is addressing the first step in this construction, that is the determination of key properties of the underlying dark matter haloes, 
in the cosmological context under study. Galaxy populations, secondary effects and cosmology forecasts will be part of future papers. 
On the longer term, we are hoping that our catalogues and gravitational lenses will be used to test new ideas that might contribute to the systematics of weak lensing signals, or to other aspects of cosmology.

One of the limitation of our simulation suite is that it does not include baryonic matter, 
hence does not model any of the baryonic physics that might influence the dark matter distribution. 
Recent work suggests that effects such as AGN feedback and supernovae winds could impact 
noticeably the matter distribution in the Universe \citep{2011MNRAS.417.2020S}. 
This might be significant for the interpretation of the lensing signal, especially at small angular scales, 
and one could imagine that future generations of simulations could implement all the effects 
we are discussing here plus the effect of baryonic feedback.

This paper is organized a follow. In Section \ref{sec:background}, we briefly review the theoretical background relevant for weak lensing studies, 
then we describe in Section  \ref{sec:num_methods} our design strategy, our N-body simulations, as well as our numerical methods 
to construct the lines-of-sight and the gravitational lenses. In Section \ref{sec:tests}, we quantify the accuracy of our simulations and of our lines-of-sight. 
We present the weak lensing estimators and their non-Gaussian uncertainty in Sections \ref{sec:estimators},
\ref{sec:estimators2} and \ref{sec:estimators3}, and conclude in Section \ref{sec:conclusion}.

\section{Theory of Weak Lensing}
\label{sec:background}


The propagation of a photon bundle emitted from a source located at an angle  {\boldmath $\beta$} and observed at  {\boldmath $\theta$} in the sky 
is characterized by a Jacobian matrix $A(\mbox{\boldmath $\theta$})$:
\begin{eqnarray}
A_{ij}(\mbox{\boldmath $\theta$}) = \frac{d\mbox{\boldmath $\beta$}}{d\mbox{\boldmath $\theta$}} = (\delta_{ij} - \Psi_{ij}(\mbox{\boldmath $\theta$}) )
\label{eq:Jacobian_matrix}
\end{eqnarray}
where the matrix $\Psi_{ij}(\mbox{\boldmath$\theta$})$ encapsulates the distortion of the two-dimensional image. 
At first order, it is determined by three components, namely a convergence $\kappa$ and two shear components $\gamma_{1}$ and $\gamma_{2}$
that combine together into a complex shear $\mbox{\boldmath $\gamma$} \equiv \gamma_{1} + i \gamma_{2}$.
At second order, an asymmetric factor $\omega$ appears in the off-diagonal elements, but it has a negligible contribution in realistic situations (see, for example, the  Appendix of \cite{Schneider98}), hence we drop it. We thus write
\begin{eqnarray}
\Psi = \left( \begin{array}{cc} \kappa +\gamma_{1} &  \gamma_{2} \\
                                                       \gamma_{2} & \kappa - \gamma_{1} \\
                                                       \end{array}\right)
\label{eq:distmatrix}
\end{eqnarray}
All of these elements are locally determined by the Newtonian potential $\Phi$ via:
\begin{eqnarray}
\kappa=\frac{\Phi_{,11}+\Phi_{,22}}{2},\gamma_1=\frac{\Phi_{,11}-\Phi_{,22}}{2}, \gamma_2=\Phi_{,12}
\label{eq:kappa_gamma}
\end{eqnarray}
where `$,i$' refers to a derivative with respect to the coordinate $i$.
For a source located at a comoving distance $\chi_{s}$, the projected distortion is computed as:
\begin{eqnarray}
\Psi_{ij} = \frac{2}{c^{2}}\int_{0}^{\chi_{s}}  \Phi_{,ij} \frac{D(\chi)D(\chi_s-\chi)}{D(\chi_s)} d\chi
\label{eq:Newton}
\end{eqnarray}
where $c$ is the speed of light. The angular diameter distance $D(\chi)$ depends on the curvature:
\begin{eqnarray}
D(\chi) =  \begin{cases}
K^{(-\frac{1}{2})} \sinh (K^{\frac{1}{2}} \chi)  & \mbox{for } \mbox{$K$ $>$ 0} \\
\chi                                                                      & \mbox{for }  \mbox{$K$ $=$ 0} \\
 -K^{(-\frac{1}{2})} \sin (-K^{\frac{1}{2}} \chi) & \mbox{for } \mbox{$K$ $<$ 0} 
\end{cases}
\end{eqnarray}
with
\begin{eqnarray}
K=\left( \frac{H_0}{c} \right)^2 (1-\Omega_m -\Omega_\Lambda)
\label{eq:k}
\end{eqnarray}
$H_{0}$ is Hubble's parameter, $\Omega_{m}$ and $\Omega_{\Lambda}$ are respectively the ratio of the mass and dark energy densities to the critical density.

The convergence field is particularly interesting theoretically since
it relates, through Poisson's equation, to the matter density contrast $\delta$:
\begin{eqnarray}
2\kappa = \nabla^2 \Phi ({\bf x})= \frac{3}{2}\Omega_{m} H_{0}^2 (1+z) \delta({\bf x})
\label{eq:gravpot}
\end{eqnarray}
with 
\begin{eqnarray}
\delta({\bf x})=\frac{\rho({\bf x})-\bar{\rho}}{\bar{\rho}}
\label{eq:delta}
\end{eqnarray}
Following standard notation, $\bar{\rho}$ is the average matter density in the Universe and  $\rho({\bf x})$ is the local density. 
In this paper, we are assuming a flat Universe, in which $D(\chi)$ takes the simplest form.
Substituting [Eq. \ref{eq:gravpot}] in [Eq. \ref{eq:Newton}], we can extract the projected convergence $\kappa$  up to a distance $\chi_s$ as:
\begin{eqnarray}
\kappa(\mbox{\boldmath $\theta$},\chi_s) \simeq  \int_0^{\chi_s} W(\chi) \delta(\chi,\mbox{\boldmath $\theta$}) d\chi 
\label{eq:kappa}
\end{eqnarray}
where ${W(\chi)}$  is defined as
\begin{eqnarray}
W(\chi)=\frac{3 \Omega_m H_0^2}{2 c^2} (1+z)g(\chi)
\label{eq:wxi}
\end{eqnarray}
and
\begin{eqnarray}
g(\chi) = \chi \int_{\chi}^{\infty} d\chi' n(\chi')\left(\frac{\chi' - \chi}{\chi'} \right)
\end{eqnarray}
In this paper,  we work under the single source plane approximation, both for illustrative purposes and to disentangle cleanly
the sources from the lenses. In this case, $g(\chi)$ reduces to $\chi(1 - \frac{\chi}{\chi_s})$ for a source plane located at $\chi_s$.
In future papers, the source distribution will be made more realistic by constructing $n(\chi)$ from observed surveys.
Finally, once we have a convergence field, we extract the shear field by solving for the gravitational potential from [Eq. \ref{eq:kappa_gamma}] \citep{1993ApJ...404..441K}\footnote{For this operation, we are working under  a flat sky approximation, 
which allows us to perform the Fourier transforms in the traditional plane wave basis, and to 
simplify the derivatives as a simple finite difference.
Also, one must be careful about the method 
used to perform this calculation, since the Universe is not periodic, while simulations usually are.
The edge effects can therefore contaminate the calculations, hence it is necessary to somehow pad the boundaries.}.

\section{Numerical Methods}
\label{sec:num_methods}

As mentioned in the introduction, gravity is a non-linear process, and the predictions from the linear theory of large scale structures are only valid on the largest scales. In the context of  weak lensing, photons trajectories are probing a broad 
dynamical range, including galactic scale structures, where the matter fields are highly non-linear.
Although higher-order perturbation theory can  describe such systems, the accuracy of the calculations are limited by the complex dynamics. 
We therefore rely on N-body simulations to generate accurate non-linear densities, through which we shoot photons rays
and extract the resulting distortion matrix.
In this section, we describe some of the considerations one must keep in mind when performing  such calculations.

\subsection{Constructing the lenses}

N-body simulations need to be optimized according to the specific measurements to be performed. 
In the current paper, we attempt to estimate the covariance matrices for a number
of weak lensing estimators, hence we need a large number of realizations.
In addition, we are interested in resolving the sub-arc minute signal, hence the simulation grid
needs to be fine.
Ideally, one would simulate the complete past light cone that connects the observer 
to the light sources all at once.
Unfortunately, for sources that extend to redshift of a few, this is not possible since the far end of the cosmological volume is at an earlier time than the close end.
It is, however, the only way one could model the  largest radial modes of a survey. 
Luckily,  it was realized that these modes contribute very little  \citep{1953ApJ...117..134L}:
the coherence scales of the largest structures significant for the signal rarely extends 
over more than a few times the size of large clusters. Simulation volumes of the order of a few hundreds of $h^{-1}$Mpc's per side
are thus generally adequate.
These simulated boxes can then be stacked to create a  line-of-sight (LOS),
inside of which photons are shot.

One can use a different simulation for each redshift box, as done by \citep{2000ApJ...537....1W},
but this method is {\small CPU} consuming, since a single LOS that extends to $z~3$ involves between 10 and 40 N-body simulations.
This is unrealistic for  covariance matrix measurements, which require hundreds of these high precisions LOS.
We opted for the now common work around developed by \cite{1998ApJ...493...10P}: 
we treat the density dumps of a given simulation as 
different sub-volumes of the same past light cone. 
To break the artificial correlation that exists across simulated volumes at different redshifts,
we perform  a rotation of each box plus a random shift of its origin.

The next step consists in calculating the photon geodesics through the large scale structures,
and to compute the cumulative deformation acquired along each trajectory.
The most accurate method would propagate the photons in a full three-dimensional volume,
with the distortion computed along the deflected trajectory.
It was shown that this is an overkill, and that calculating the distortion only at the mid-planes
of the box provides weak lensing fields  that differ by no more than $0.1$ per cent with the full three dimensional treatment \citep{2003ApJ...592..699V}.
Moreover, it was shown at the same time that this simplification has nearly  indistinguishable effects on the two- and three-point functions.
There is thus no need to store the full dark matter density field, a significant advantage when working with hundreds 
of high resolution N-body simulations.
For the same reason, we did not adopted the three-dimensional lensing calculation proposed by \cite{2011MNRAS.414.2235K},
because it requires storing the particle catalogs of the full past light cones.
Many authors have since opted for such ray tracing or line-of-sight integration techniques 
\citep{2000ApJ...530..547J, 2007MNRAS.379.1507F, 2009A&A...499...31H}.

The first step in this approach consists in collapsing the cosmological sub-volumes into their mid-planes, and 
calculating the geodesics on specified angular locations, or pixels, on these thin lenses.  
In the weak lensing regime, these  trajectories are close to straight lines, such that Born's approximation is very accurate \citep{Schneider98, 2004APh....22...19W}. 
In this paper, we therefore opt for a line-of-sight integration along the unperturbed photon paths. 
We nevertheless stored the full simulated  lenses, allowing future analysis to test how much ray-tracing or post-Born calculations \citep{Schneider98, 2010A&A...523A..28K} affect the results.

N-body computations in this setting are fast, but not optimal: the interpolation becomes very strong at low redshift, 
thus increasing the impact of the simulation softening length. More over, a large portion of the simulated volume is left unused:  
the past light cone,  shaped like a truncated pyramid, is extracted from a cuboid. 
One way to improve on these two effects is to reduce the size of the simulation box for low redshift dumps,
as done in  \cite{2000ApJ...537....1W} for example, who used six different box sizes to reach sources at $z=1$.
That involved running six times more independent simulations, a price that is not always affordable.
In the current work, it would have been too expensive to run that many, 
but two  distinct volume sizes offers a reasonable trade off.

Since there is inevitably some wasted cosmological volume, we could in principle re-shuffle the projections axis and the origin to create about ten time more LOS from the same simulations, following \cite{2000ApJ...530..547J}\footnote{We have used some of this wasted volume on occasions to extract
lenses at low redshifts for some simulations that could not run until the end (reaching hard walltime limit, computer cluster downtime, growth of
ultra dense regions that unbalance the work load, etc.). 
In that process, we made sure that no volume was used by more than one LOS
by constraining the random shifts to point to virgin domains. There is therefore no extra correlation induced.}.
 However, this would inevitably produce a small amount of extra correlation between different realizations, which would propagate and contaminate the covariance matrix with extra non-Gaussian features. We opted out from this option, however we saved the full mid-planes, for usages like this to be available for future analyses requiring even larger statistics.

\begin{figure}
  \begin{center}
\epsfig{file= 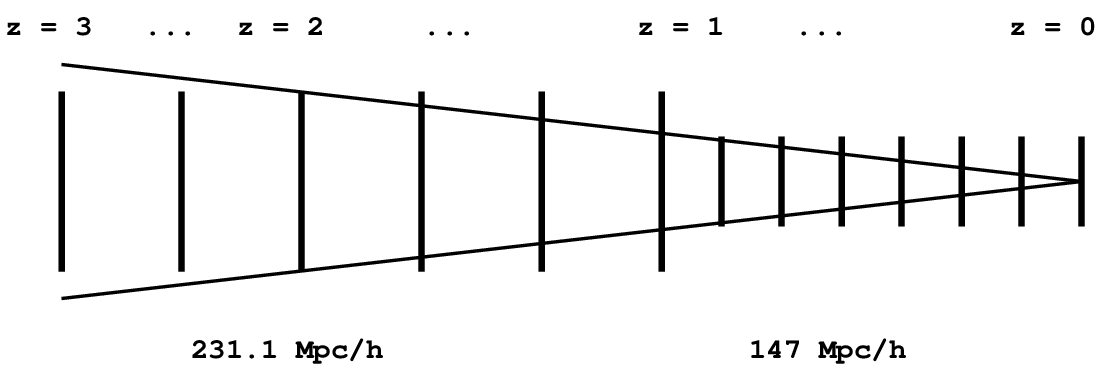,width=0.4\textwidth,height=0.2\textwidth}

  \caption{Geometry of the lines-of-sight. The global simulated volume consists of two adjacent rectangular prisms, 
  collapsed as a series of thin lenses. As explained in the text, high redshift volumes are larger,
   but the number of simulated grid cells and pixels are kept constant.
   The observer sits at $z=0$;  the junction between the small (lower-$z$) and large (higher-$z$)
  simulation boxes occurs at $z=1$;  the past light cone escapes the simulated volume beyond $z = 2$, where we exploit the periodicity 
  of the boundary condition to populate the edges; we store lenses and haloes up to $z = 3$.
   \label{fig:geometry}}
  \end{center}
\end{figure}

As mentioned in the introduction, this work is meant to outperform the dynamical range of previous weak lensing simulations: we need sub-arc minute precision,
 with a field of view of a few degrees per side.
We design our LOS such that each pixel has an opening angle of $0.21$  arc minute on each side,
with $n_{pix} = 1024^{2}$ pixels in total, for a total opening angle of $3.58^{o}$ per side.
Moreover, we constructed the survey geometry such that rays shot at $z=0$ will reach the edge of the small simulation box at  $z=1$.
This uniquely specifies the box size of our low-redshift simulations:
with our choice of cosmology, we get $L=147.0$ $h^{-1}$Mpc per side.
Rays then enter higher redshifts  boxes, which have a larger size.
That way, no unperturbed photon paths actually escape the volume yet.
Our second requirement is that the surface of the past light cone reaches the edge of the larger box at $z=2$.
This yields a  comoving side of $L = 231.1$ $h^{-1}$Mpc.
Some of the outer rays eventually leave the simulated volume at redshifts larger than $2.0$, 
in which case we enforce the periodicity of the simulations box(see Fig. \ref{fig:geometry}).
This situation applies only to the last four lenses, hence the total amount of repeated structures is very small. 
This is even further suppressed by the lensing kernel, which favours redshifts closer to $z=1-1.5$, and by the fact such high
redshifts have fewer galaxies to start with.
We could have avoided this `leakage' by choosing a larger volume for the high redshift boxes,
but the resolution would have been penalized in a critical region.

\subsection{N-Body simulations}
\label{subsec:nbody}

The N-body simulations are produced by {\small CUBEP3M}, an improved version of {\small PMFAST} \citep{2005NewA...10..393M} that 
is both {\small MPI} and {\small OPENMP} parallel, memory local and also allows for particle-particle (pp) interaction
at the sub-grid level. 
$1024^{3}$ particles are placed on a $2048^{3}$ grid, and have their initial displacements and velocities 
calculated from the Zel'dovich approximation \citep{1970A&A.....5...84Z,1989RvMP...61..185S}. The transfer function that enters this calculation was obtained 
from {\small CAMB} \citep{1996ApJ...469..437S}. We used the WMAP5 cosmology \citep{2009ApJS..180..330K} in our simulations and theoretical predictions : $\Omega_{\Lambda} = 0.721$, 
$\Omega_{m} = 0.279$, $\Omega_{b} = 0.046$, $n_{s} = 0.96$, $\sigma_{8} = 0.817$ and $h = 0.701$.
This cosmology and the simulation volumes discussed above completely specify the mass of the particles: in the large (small) boxes, we have  
$m_{p} =1.2759 \times 10^{9}$  ($3.2837 \times 10^{8}$)  $M_{\odot}$. 
Also, the comoving sub-grid softening lengths $r_{soft}$ are of $112.8$ and $71.8$ $h^{-1}$kpc respectively\footnote{
$r_{soft}$  is defined to be a tenth of a grid cell, and is enforced by a sharp cutoff in the force of gravity for particle pairs separated by smaller distances.}.
At the transition between the two volumes, this change in mass 
affects the mass function of the halo catalogues, since the smallest collapsed structures  do not have the same physical mass.
However, the weak lensing signals rely only on contrasts in grid projections, which keep no trace of sub-grid level objects.

The initial redshifts are selected such as to optimize both the run time and the accuracy of the N-body code. 
These are chosen to be $z_{i} = 40.0$ and $200.0$ for the large and small boxes respectively.
The reason for selecting different starting redshifts resides in the fact that the smaller volumes are probing smaller scales, hence they need to start earlier, at a time where the Nyquist frequency of the grid is well in the linear regime. Each simulation is then evolved with {\small CUBEP3M}
on 8 nodes of the {\small TCS} (Tightly Coupled System) supercomputer  at the SciNet HPC Consortium \citep{SciNet} --
to which system we dedicate the name of the simulation suite. 
The lens redshifts, $z_{l}$, are found by breaking the comoving distance between  $z=0.0$ and $z=1.0$ into cubes of $L=147.0$ $h^{-1}$Mpc per side
(and that between $z=1.0$ and $z=3.0$  into cubes of $L=231.1$ $h^{-1}$Mpc per side), and solving for the redshift at each mid-planes. 
The resulting redshifts are presented in Table \ref{table:redshifts}.
When the simulations reach these redshifts, the dark matter particles are placed on to a $2048^{3}$ grid following the `cloud-in-cell' interpolation scheme \citep{1981csup.book.....H}, and the grid is then collapsed into a slab along a randomly selected axis.

\begin{table}
  \begin{center}
\caption{Redshifts of the lenses. The projections for $z_l>1.0$ are produced with  $L=231.1$  $h^{-1}$Mpc simulations, while those for lower $z_l$ are obtained from an independent set of simulations with  $L = 147.0$ $h^{-1}$Mpc. 
\label{table:redshifts}}
\begin{tabular}{|c|c|c|c|c|c|c|} 
\hline
3.004&
2.691&
2.411&
2.159&
1.933&
1.728&
1.542\\
\hline
1.371&
1.215&
1.071&
0.961&
0.881&
0.804&
0.730\\
\hline
0.659&
0.591&
0.526&
0.463&
0.402&
0.344&
0.287\\
\hline
0.232&
0.178&
0.126&
0.075&
0.025& & \\ \hline
\end{tabular}
\end{center}
\end{table}

With this configuration, we solve [Eq. \ref{eq:kappa}] numerically for each pixel.
We convert the $\chi$ integral into a discrete sum at the lens locations $\chi(z_{l})$. 
The infinitesimal element $d\chi$ becomes $dL/n_{grid}$, where $n_{grid} = 2048$
and $L = 147.0$ or $231.1$ $h^{-1}$Mpc, depending on the redshift of the lens.
Under the single source plane approximation, we can thus write the convergence field as
\begin{eqnarray}
   \kappa({\bf x})= \frac{3 H_{o} ^{2} \Omega_{m} } {2 c^{2} } \sum_{z_{l}}^{z_{s}} \tilde{\delta}({\bf x}) (1+z_{l}) \chi(z_{l})  (1-\chi(z_{l})/\chi(z_{s})) d\chi 
   \label{eq:kappa_discrete}
\end{eqnarray}
 where $\tilde{\delta}({\bf x})$ is the two-dimensional density contrast on the mid-plane\footnote{
To avoid edge effects when computing the shear fields, we perform the Fourier transforms on the full periodic slabs, i.e. {\it before} the interpolation on to the lenses. 
As explained at the very end of section \ref{sec:background}, one would otherwise have to zero pad the convergence fields.}

\section{Testing the simulations}
\label{sec:tests}

In this section, we quantify the resolution and accuracy of the N-body simulations.
 We  measure the power spectrum of the three-dimensional density fields -- i.e. before collapsing onto mid-planes, 
 and before interpolating on to the pixel locations of the past light cone -- and 
  extract the angular power spectrum of the lines-of-sight. In both cases, we compare
  our results to non-linear predictions and identify the limitations of our calculations.
 

\subsection{Dark matter density power spectrum}
\label{subsec:dm}

\begin{figure}
\begin{center}
\epsfig{file= 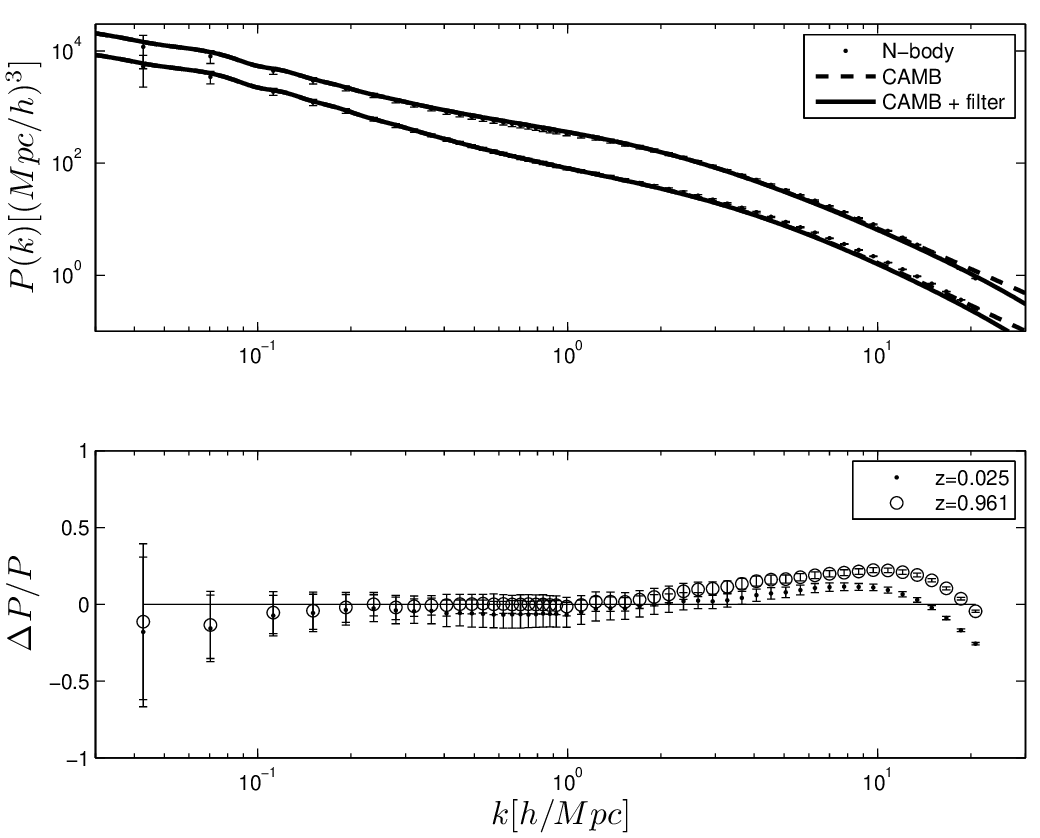,width=0.49\textwidth,height=0.49\textwidth}
\caption{({\it top}:) Power spectrum of $185$ N-body simulations, at redshifts of $0.961$ (bottom curve) and $0.025$ (top curve). 
The solid and dashed lines are the non-linear predictions, with and without the Gaussian filter respectively. 
The error bars shown here are the standard deviation over our sampling. We observe a slight over estimate of power in the simulations for scales
smaller than $k = 3.0 h \mbox{Mpc}^{-1}$. 
This is caused by a known loss of accuracy in the predictions, as one progresses deep in the non-linear regime.
At the same time, resolution effects are affecting these scales, with a turn over at $k \sim 10.0 h \mbox{Mpc}^{-1}$.
This regime is thus not to be trusted in terms of accuracy.
({\it bottom}:) Fractional error between the simulations and the non-linear
predictions. \label{fig:dm_ps}}
\end{center}
\end{figure}

The power spectrum of the matter density $P(k)$ is a fast and informative test of the quality of the simulations.
It probes the growth of structures at all scales available within the volume,
and in comparison with a reliable theoretical model, informs us about both the accuracy and the resolution limits of the simulations.
For a given density contrast $\delta({\bf x})$, the power spectrum can be calculated from its Fourier transform $\delta({\bf k})$ as:
\begin{eqnarray}
    \langle  \delta({\bf k}) \delta({\bf k'}) \rangle  = (2\pi)^{3} \delta_{D}({\bf k}-{\bf k'})P({\bf k}) 
\end{eqnarray}
where the angle brackets refer to a volume average, and the Dirac delta function selects identical Fourier modes.
 We present  in Fig. \ref{fig:dm_ps} the power spectrum for our 185 simulations at two redshifts, $z = 0.961$ and $z=0.025$.
When compared with the theoretical predictions from {\small CAMB} \citep{Lewis:1999bs}, 
we observe that the lower redshift power is a few per cent lower for $0.4 < k < 1.0 $ $h \mbox{Mpc}^{-1}$,
and that the low (high) redshift over estimate the power by about ten (twenty) per cent for  $ k > 1.0 $ $h \mbox{Mpc}^{-1}$.
These discrepancies can be caused by a number of effects, from finite box size to residual uncertainty in the numerical integration, 
and inevitably propagate in the calculations of past light cone. 
However, deviations from {\small CAMB} are observed in most N-body codes -- \cite{2010MNRAS.408..300G} measured up to 50 per cent deviation in the Millenium simulation --, and do not affect qualitatively 
the final results, as long as internal consistency is preserved\footnote{The version of {\small CAMB} that was used in our calculations
 does not incorporate recent corrections that were made in December 2011}.

In N-body codes, resolution limits are mainly determined by the softening length in the gravitational force, and typically cause an abrupt drop in the observed power spectrum at small scales. This drop of power can be modelled, following \citep{2003ApJ...592..699V}, by a Gaussian filtering of the form $\mbox{exp}[-k^{2}\sigma_{g}^{2}]$ in the power spectrum, where $\sigma_g = 0.155 L/N_{grid}$. In our simulations, we observe that the structures seem to be well modelled down to $k = 10.0  h \mbox{Mpc}^{-1}$, which corresponds to a comoving length of about $630$ $h^{-1}$ kpc. 
It is possible to obtain a rough estimate of the impact of this resolution limit on the weak lensing angular power spectrum as follow:
we know that the redshift that dominates the lensing kernel is about $z\sim 1.0$, hence that at this distance, the Gaussian
filter $1\sigma_{g}$ subtends an angle of  $\theta_{soft} = 0.148$ arcsec.
Of course, the signal is sensitive to structure much closer, where the softening angle becomes much higher.
 This technique also depends on the details of the N-body code, and a more accurate estimate of the resolution limit
is found from a comparison of the simulated angular power spectrum with a reliable non-linear model
(see section \ref{subsec:Cl}).

When it comes to measuring the uncertainty on the matter power spectrum,
most data analyses worked in the framework of linear theory of structure formation.
Notably, this presumes that different Fourier modes of the matter density grow independently, 
such that the error bars on the power spectrum are well described by Gaussian statistics. 
For non-linear processes, however,  the phases of different Fourier modes start to couple together \citep{1999MNRAS.308.1179M,2000Natur.406..376C, 2002MNRAS.337..488C}, 
therefore higher order statistics, i.e. bispectrum, trispectrum, are then needed in order to completely
characterize the field. Theoretical calculations can describe these quantities with a reasonable accuracy, at least in the trans-linear regime,
but N-body simulations provide the most accurate estimates at all scales resolved.
These simulations are primarily used to extract the non-Gaussian covariance matrices of various weak lensing estimators, hence it is essential to pin down the sources non-Gaussian features, and to monitor how these propagate in our calculations.
Our plan is to organize the final lensing estimators in about ten angular bins, 
hence 185 simulations  are enough to ensure convergence on each element of the covariance matrices.

\begin{figure}
\begin{center}
\epsfig{file= 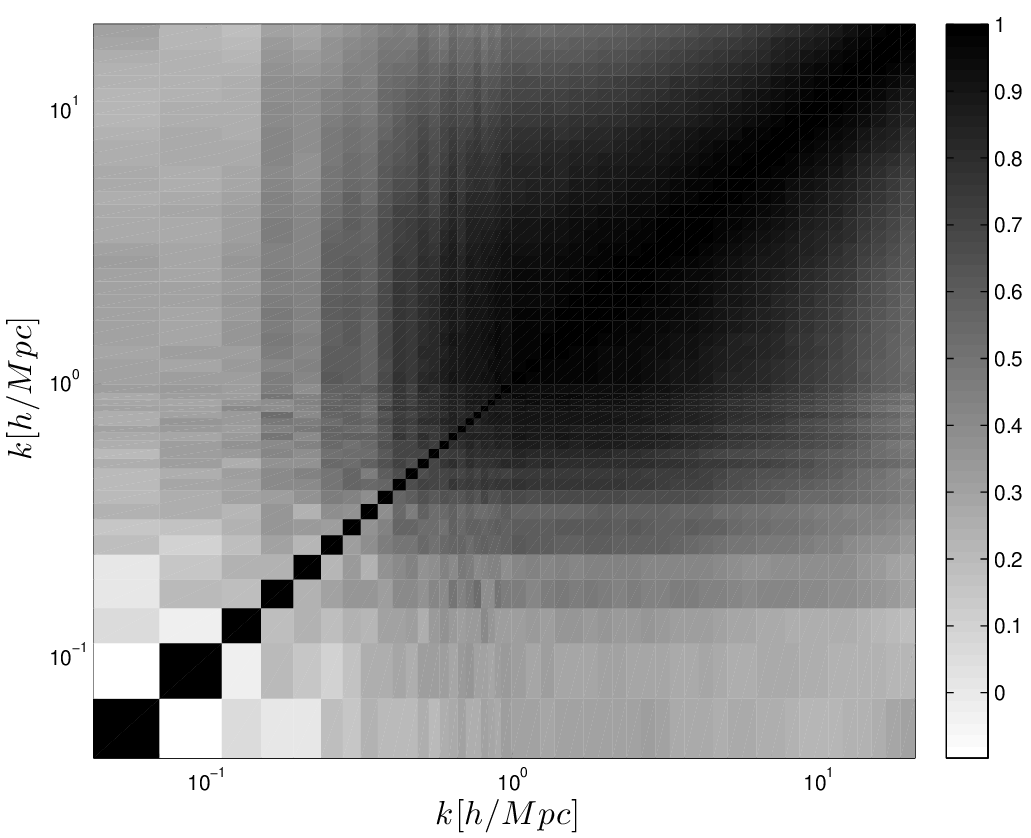,width=0.49\textwidth,height=0.49\textwidth}
\caption{Cross-correlation coefficient matrix of the density power spectrum, measured from of $185$ N-body simulations, 
at redshift of $0.961$. Modes at $k\sim 0.5 h \mbox{Mpc}^{-1}$, corresponding to $\theta \sim 18.35$ arcmin, are more than 40 per cent correlated. 
\label{fig:dm_cov2}}
\end{center}
\end{figure}

The power spectrum covariance matrix is defined as 
\begin{eqnarray}
   C(k,k') = \langle P(k) -  \bar{P}(k) \rangle \langle P(k') -  \bar{P}(k') \rangle 
\end{eqnarray}
where the over-bar refers to the best estimate of the mean. 
The amount of correlation between different scales is better visualized
with the cross-correlation coefficient matrix, which is obtained from $C(k,k')$ via
\begin{eqnarray}
   \rho(k,k') = \frac{C(k,k')}{\sqrt{C(k,k)C(k',k')}}
\end{eqnarray}
and is shown for $z=0.961$ in  Fig. \ref{fig:dm_cov2}. 
We see that it is almost diagonal for large scales (low $k$), while measurements become correlated as we progress towards smaller scales (higher $k$).
At $k\sim 0.5 h \mbox{Mpc}^{-1}$, for instance, the Fourier modes are more that 40 per cent correlated. At this redshift, these correlated scales correspond to angles smaller than $\theta \sim 18.35$ arcmin on the sky, or to $\ell > 1180$.
We stress that this correlation is due to the sole effect of non-linear dynamics, 
is thus intrinsic to the density fields.
Recent results have shown that neglecting either the 
non-Gaussian nature of the uncertainty can significantly 
underestimate the error on the power spectrum, even at scales traditionally considered as linear  
\citep{2011arXiv1106.5548N,2011arXiv1109.5746H}.


\subsection{Weak lensing power spectrum}
\label{subsec:Cl}

\begin{figure}
\begin{center}
\epsfig{file= 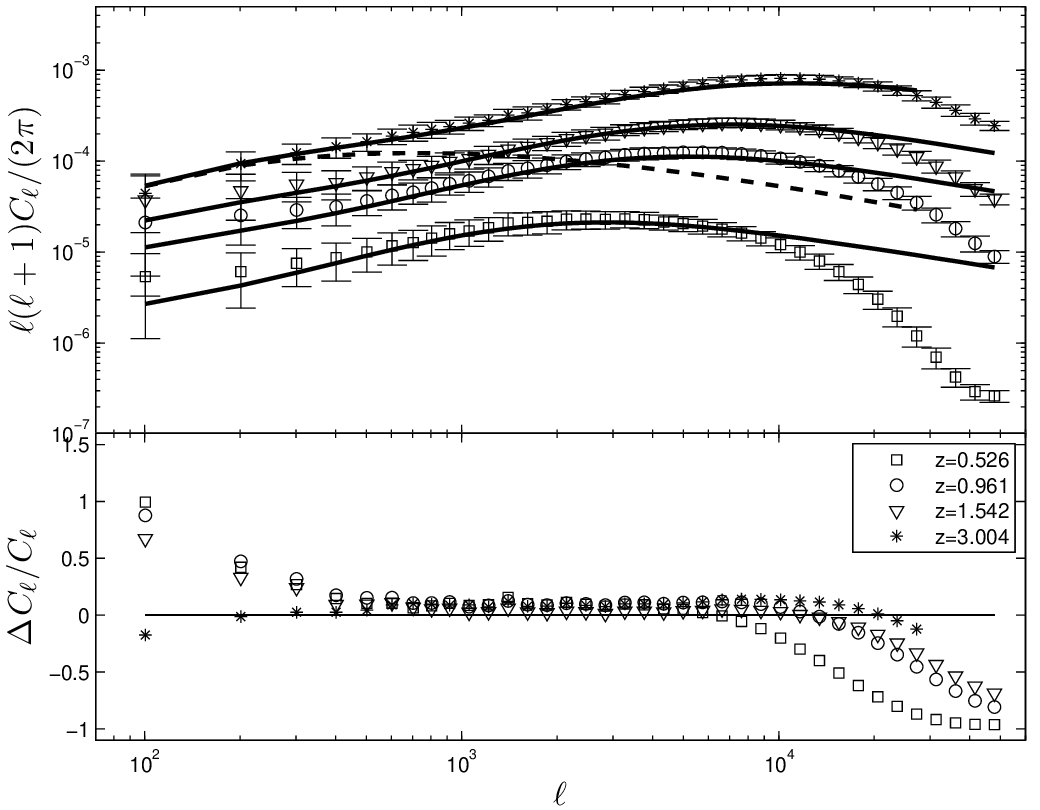,width=0.49\textwidth,height=0.49\textwidth}
\caption{({\it top}:) Convergence power spectrum, measured from $185$ N-body simulations,
where the source redshift distribution is a Dirac delta function at $z\sim3.0,1.5,1.0$ and $0.5$ (top to bottom symbols).
The solid lines correspond to the non-linear predictions, which are calculated with a modification of 
the {\small CAMB} package  \citep{1996ApJ...469..437S}, where the sources and lenses  have been placed according to the survey depth. The linear predictions at $z_s=3$ are represented by the  dashed line, and the error bars are the 1$\sigma$ standard deviation over our sampling. 
We observe a slight over-estimate of power in the simulations for $z=3$ and $\ell > 1000$ compared to non-linear predictions (solid line),
and a more important bias for lower redshifts. 
This is caused by an underestimate of the power spectrum in the theoretical predictions,  which is also visible in the smallest scales of the three dimensional dark matter power spectrum (i.e. Fig. \ref{fig:dm_ps}). Similar trends are observed in the Coyote Universe  and {\small SUNGLASS} simulation suites. 
The low-$\ell$ power seems also to be in excess in the simulations, however predictions are still with the error bars. ({\it bottom}:) Fractional error between the simulations and the non-linear predictions. \label{fig:Cl_kappa}}
\end{center}
\end{figure}

In order to  understand the resolution of our lensing maps, we measure the angular power spectrum of the $\kappa(\mbox{\boldmath $\theta$})$ fields,
and compare the results with the non-linear predictions. 
These are obtained from a simple modification of the {\small CAMB} package \citep{1996ApJ...469..437S},
in which the redshift window function has been adjusted to match that of the current survey geometry. It  uses Limber's approximation to express the convergence power spectrum as 
an integral over $P(k)$. The  power spectrum of the convergence field  is defined as:
\begin{eqnarray}
\langle \kappa(\mbox{\boldmath $\ell$})\kappa(\mbox{\boldmath $\ell'$}) \rangle=(2 \pi)^2 \delta_D(\mbox{\boldmath $\ell$}+\mbox{\boldmath $\ell'$})C_{\ell}
\label{eq:kappacorr}
\end{eqnarray}
where $\mbox{\boldmath $\ell$}$ is the Fourier component corresponding to the real space vector $\mbox{\boldmath $\theta$}$.
and where, again, the angle brackets refer to an angle average.
The convergence power spectrum, estimated from our simulations, 
is shown in Fig. \ref{fig:Cl_kappa} where the error bars are the $1\sigma$ standard deviation on the sampling. When compared with the non-linear model, we find a good agreement
in lower multipoles, while the theoretical predictions slightly underestimate the power for $\ell > 1000$, consistent with the observations of \cite{2009A&A...499...31H}.
The strong departure at $\ell \sim 30000$ is caused by limitations in the resolution, 
and corresponds to an angle of about  $0.7$ arcmin.

\begin{figure}
\begin{center}
\epsfig{file= 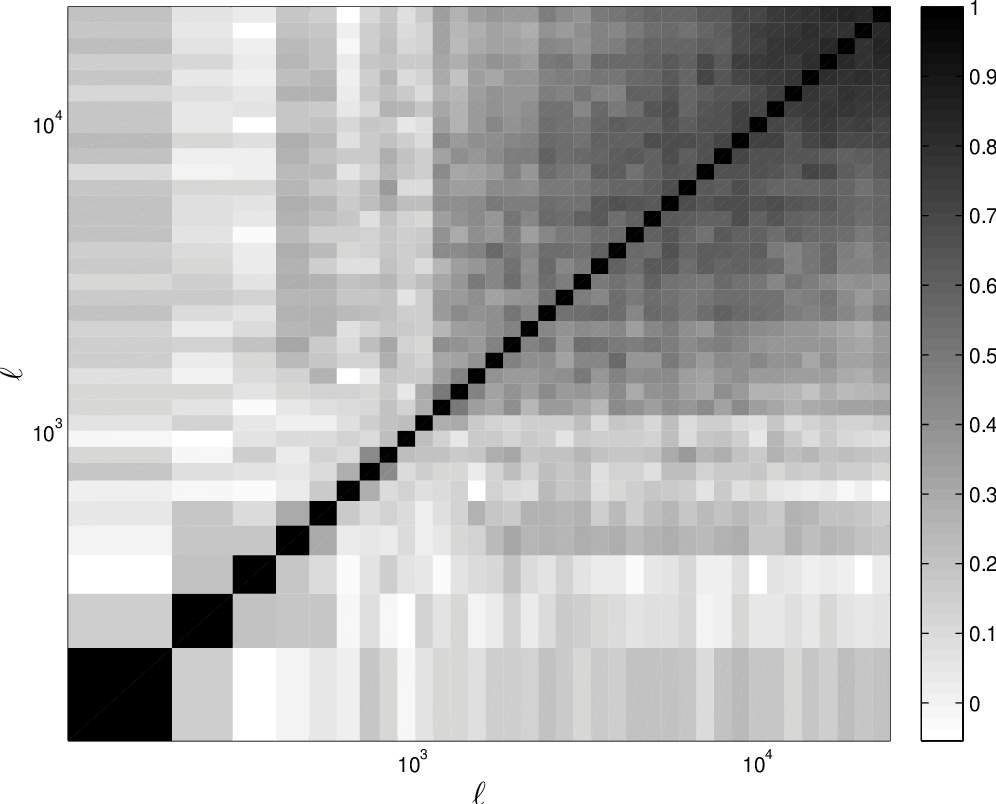,width=0.49\textwidth,height=0.49\textwidth}
\caption{Cross-correlation coefficient matrix of the (dimension-full) convergence power spectrum, 
measured from $185$ LOS. We observe a strong correlation for $\ell$ of a few thousand,
consistent with the findings of \protect \cite{2009arXiv0905.0501D}.\label{fig:kappa_cov}}
\end{center}
\end{figure}

As mentioned earlier, the smallest angles of weak lensing observations are probing the non-linear regime of the underlying density field, and it is known that the statistics describing the uncertainty
on $C_{\ell}$ are non-Gaussian. As a matter of fact,  \cite{2009arXiv0905.0501D}
have demonstrated that the non-Gaussian features in the weak lensing fields
lead to a significant loss of constraining power on the dark energy equation of state\footnote{Following the jargon developed in the Dark Energy Task Force \citep{2006astro.ph..9591A}, a lower `figure-of-merit' corresponds to a larger error about the equation of state.}
 (see figure 6 in their paper). For instance, at $\ell = 1000$, the figure-of-merit differs by 50 per cent when compared 
to Gaussian calculations, and the difference is even larger when including higher multipoles.
Although most of the departures from Gaussianity in the data are currently lost in the observation noise,
future lensing surveys are expected to improve enough  on statistics and systematics such that non-Gaussian features
will become significant. The non-linear dynamics effectively correlate the error bars in small scales,
as seen  in Fig. \ref{fig:kappa_cov}.
As expected, we observe that all the multipoles with $\ell > 1000$ are more than $40$ per cent correlated.

\subsection{Halo Catalogues}
\label{subsec:halo}

This section briefly describes how the halo catalogues are created, 
and presents a few of their statistical properties.
We recall that one of our main objective is to construct mock galaxy catalogues
on which many secondary effects will be quantified.
As mentioned before, we do not attempt to populate the haloes in this paper, since this is a challenge on its own,
and we wish to factor out this problem for now. In the future, though, one could follow the strategy of \cite{2006MNRAS.371..750H}, and populate the haloes with galaxies under the conditional luminosity function of \cite{2005ApJ...627L..89C}, then assign ellipticity following the elliptical or spiral  model \citep{2000MNRAS.319..649H, 2004MNRAS.347..895H}. Alternatively, one could use the {\small GALICS} \citep{2003MNRAS.343...75H} and {\small MOMAF} \citep{2005MNRAS.360..159B} pipelines to create mock galaxy catalogues directly  from our halo catalogues, following the prescription 
described in \cite{2007MNRAS.379.1507F}. Our plan is to incorporate, for the first time and in a systematic way, 
the prescriptions for intrinsic alignment, source clustering, ellipticity-shear correlation, etc. 
\citep{2007MNRAS.375....2D, 2010MNRAS.402.2127S} all at once. This is crucial in order to interpret correctly the signal from the data,
which contains all these contributions.

In this work, haloes are constructed from the matter density fields with a spherical overdensity search algorithm \citep{1996MNRAS.281..716C}. 
The first step is to assign the dark matter particles on a $2048^3$ grid with the Nearest Grid Point 
scheme \citep{1981csup.book.....H}, and identify local density  maxima. The halo finder then ranks these candidates in decreasing order of peak height, and for those which are above an inspection  threshold value, it grows a spherical volume centred on the peak, computing for each shell the integrated overdensity until it drops under the predicted critical value. The haloes that are analyzed first are then removed from the density field, in preparation for the inspection of lower mass candidates. This prevents
particles from contributing multiple times, but at the same time limits the resolution on sub-structures of the largest haloes. 
This is a mild cost for the purpose of these catalogues, which are populated with low multiplicity of galaxies,
and thus depend rather weakly on the sub-halo structures.
Finally, for each halo, we measure the mass, the centre-of-mass (CM) and peak positions, the CM velocity,
the velocity dispersion, the angular momentum in the  CM frame, and the inertia matrix $\sigma_{ij}$, 
which allows us to use population algorithms that outperform those that depend solely on the halo mass.
 Although the inertia matrix is biased by the fact that we are only searching for spherical regions,
 we still recover significant information about the shape and orientation.

\begin{figure}
\begin{center}
\epsfig{file= 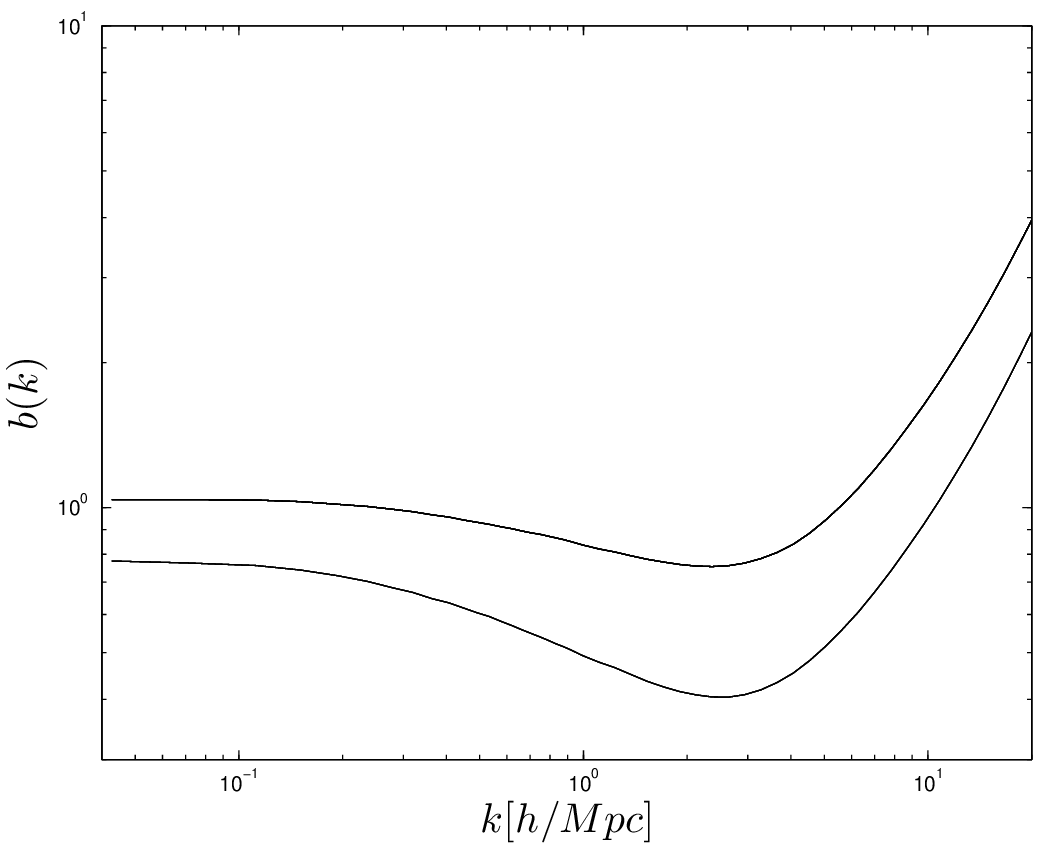,width=0.49\textwidth,height=0.49\textwidth}
\caption{Halo bias, for $z=0.025$ (bottom curve) and $z=0.961$ (top curve). 
These results are in good agreement with the results of \protect \cite{2011arXiv1107.4772I},
even though the direct comparison is complicated by the fact that the bias is dependent on both the redshift and the mass bins. \label{fig:halo_bias}}
\end{center}
\end{figure}

\begin{figure}
\begin{center}
\epsfig{file= 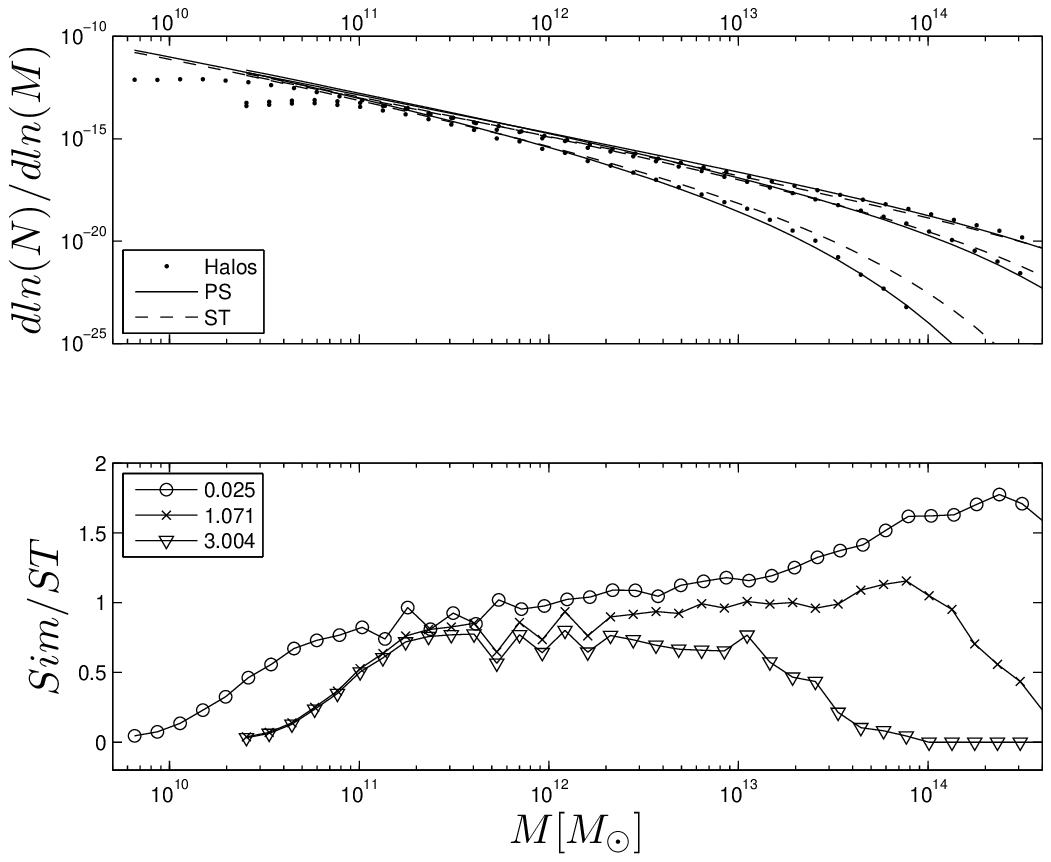,width=0.49\textwidth,height=0.49\textwidth}
\caption{ ({\it top}:) Halo mass function, compared to predictions, for redshift 0.025, 1.071 and 3.004 (top to bottom curves). 
({\it bottom}:) Ratio of the mass function to the theoretical predictions of Sheth-Tormen. \label{fig:halo_mass}}
\end{center}
\end{figure}

In all the plots of this section, we present properties of the haloes that populate the full simulation box,
even though, in the final mock catalogues, we keep only those that sit inside the past light cone. 
We apply the same coordinate rotation and random shifting of the origin that was performed on the lenses, such that the halo catalogues and the lenses trace the same underlying density field. 
To quantify the accuracy of the halo catalogues, we first extract the power spectrum of the distribution,
and compare the results with the measurements from dark matter particles by computing the 
halo bias, defined as $b(k) = \sqrt{P_{halo}(k) / P_{particle}(k)}$. From Fig. \ref{fig:halo_bias}, 
we observe that both the shape and redshift dependence agree generally well with the results from \cite{2011arXiv1107.4772I}. 
A direct comparison is complicated, however, since the bias is both redshift and mass dependent.
 We observe in Fig. \ref{fig:halo_mass} that at $z = 1.071$, the halo mass function is in good agreement with both  \cite{1974ApJ...187..425P} and  \cite{2002MNRAS.329...61S} in the range $1 \times 10^{11} - 2 \times 10^{14} M_{\odot}$.
   Higher redshifts seem to better fit Press-Schechter in the same dynamical range (bottom-most curve in the upper panel)    
   while lower redshifts are better described by the Sheth-Tormen model. These are on the high side, however,
   and reach a positive bias of about  $50$ per cent for $M > 10^{14} M_{\odot}$.

\section{Weak Lensing with 2-Point Correlation Functions}
\label{sec:estimators}

Detection of a robust weak lensing signal from the data is a challenging task in itself for many reasons. The number density of galaxies detected, with their shape resolved, needs to be high,
and many secondary effects, mentioned in the Introduction of this paper, contaminate the signal and need to be either filtered out or controlled.
Generally, different statistical estimators and filtering techniques  are sensitive to different scales, systematics and secondary effects,
and their measurements  correlate in a unique way. 
It was recently shown in \cite{Vafaei10} that the optimal approach for measurements involving the cosmic shear and convergence
depends on the survey geometry and on the cosmological parameters investigated.
For instance, the shear 2-point correlation function minimizes the correlation across different angles, while mass aperture window statistics are more sensitive to smaller scales, hence 
are better suited for surveys of limited coverage  \citep{Schneider98}. 

Understanding the non-Gaussian aspects of these estimators is the 
goal of the next three sections. For each of them, we first give a short description, 
then present  their signal and associated non-Gaussian uncertainty.
The current section covers estimates based on the 2-point functions, section \ref{sec:estimators2} discusses window-integrated estimators, and section  \ref{sec:estimators3} describes
alternative statistics based exclusively on the convergence maps.
In all cases, the theoretical predictions follow the prescriptions of \cite{2001MNRAS.322..918V},
which are third-order calculations based on perturbation theory.

The 2- and 3-point functions of the observed lensing field are known to provide a wealth of information about the underlying density field.  In this paper, we constructed $185$ independent shear and convergence maps from our N-body simulations.  As described in section 
\ref{sec:background}, these maps are extracted from the projected density of the grid,
hence information about individual particles is lost. To mimic the actual detection from a galaxy survey, we Poisson sample each of the maps with $100 000$ random points and construct mock catalogues, from which we extract the 2-point correlation function measurements.
The positions are purely random within the 12.84 deg$^2$ patches, and the values at each point 
are interpolated from the simulated grids.  This completely bypasses more realistic galaxy population algorithm, which will be addressed in future work.


\subsection{Shear}

In current weak lensing analyses, one of the strongest signal comes from a measurement of the
2-point correlation function in the shear of galaxy, which is defined as: 
\begin{eqnarray}
\xi_{ij}(\theta) \equiv \langle \gamma_i (\mbox{\boldmath $\theta'$}) \gamma_j(\mbox{\boldmath $\theta$}+\mbox{\boldmath $\theta'$}) \rangle
\end{eqnarray}
where $i$ and $j$ refer to a pair of galaxies separated by angle $\theta = |\mbox{\boldmath $\theta$}|$.
In the absence of gravitational lenses, $\xi_{ij}(\theta)$ averages out to zero, hence a positive signal
indicates a detection of cosmic shear.
The intrinsic distortion produced by a single massive object is exclusively aligned in the tangential direction around its centre of mass. 
It is therefore natural to consider a coordinate system that is local for each galaxy pairs, 
in which the {\it tangential} and {\it rotated}  axes $(t,r)$ are defined as the direction perpendicular and parallel to the line joining them, respectively.
The new components of the complex shear are
written as $\mbox{\boldmath $\gamma$}=\gamma_t + i \gamma_r$.
In fact, many of the weak lensing estimators  
can be simplified when expressed as a function of these.  

The corresponding correlation functions  $\xi_{tt}$ and  $\xi_{rr}$ are defined as the weighted average of the tangential and 
rotated shears for pairs of galaxies separated by an angle $\theta$=$|x_i-x_j|$, namely:
\begin{eqnarray}
\xi_{tt}(\theta)=\frac{\Sigma w_i w_j \gamma_t(x_i)\gamma_t(x_j)}{\Sigma w_i w_j}
\label{eq:xitt}
\end{eqnarray}
\begin{eqnarray}
\xi_{rr}(\theta)=\frac{\Sigma w_i w_j \gamma_r(x_i)\gamma_r(x_j)}{\Sigma w_i w_j}
\label{eq:xirr}
\end{eqnarray}
where the weights $w_i$ quantify how well the shear is measured on the object $i$.
A convenient linear combination of the tangential and rotated shears :
\begin{eqnarray}
\xi_\pm (\theta) = \xi_{tt} \pm \xi_{rr}
\label{eq:xipm}
\end{eqnarray}
is particularly useful since it is directly related to the convergence power spectrum:
\begin{eqnarray}
\xi_+(\theta)=\int_0^\infty \frac{d \ell}{2 \pi} \ell C_{\ell} J_0 (\ell \theta)
\label{eq:xi+}
\end{eqnarray}
\begin{eqnarray}
\xi_-(\theta)=\int_0^\infty \frac{d \ell}{2 \pi} \ell C_{\ell} J_4 (\ell \theta)
\label{eq:xi-}
\end{eqnarray}
where $J_n(x)$  is the $n$th order first kind Bessel function.
Hence measurements of $\xi_{rr,tt}$ give a direct handle on cosmological parameters that depend on  $C_{\ell}$.

\begin{figure*}
\begin{center}
\includegraphics[scale=0.70]{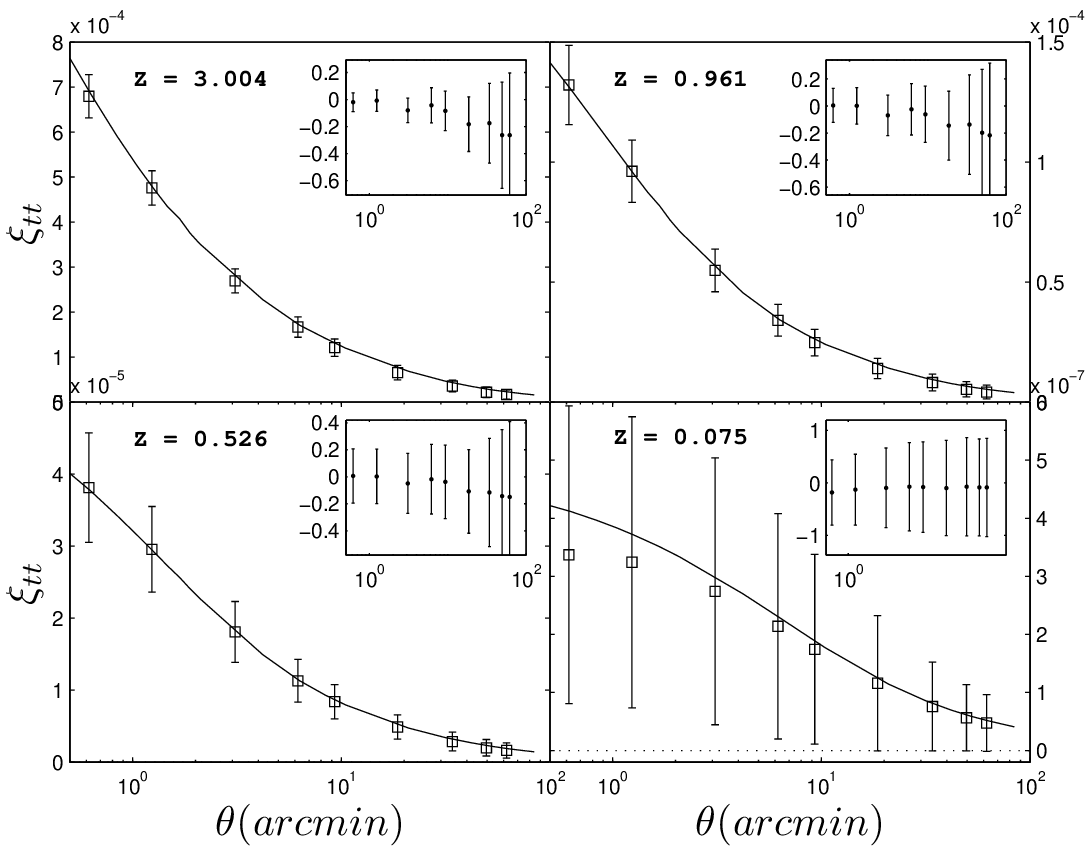}
\end{center}
\caption{Shear correlation function $\xi_{tt}$, computed from Poisson sampling the simulated shear maps
in the local $(tt,rr)$ coordinates, as described in the text. 
The solid line shows non-linear theoretical predictions on the mean of the estimators. 
Theoretical predictions for Figs. 6-17 were obtained with third order expansion in perturbation theory, as described by   \citep{2001MNRAS.322..918V}. In each of these figures,  top left to down right panels represent $z=3.004, 0.961, 0.526$ and $0.025$ respectively, and the inset
is the fractional error between the predictions and the simulations. In this figure, the fractional error
between the mean and the theory is around $10$ per cent, although the agreement is fully consistent with $1\sigma$ error bars.    
\label{fig:xitt}  }
\end{figure*}
\begin{figure*}
\begin{center}
\includegraphics[scale=0.70]{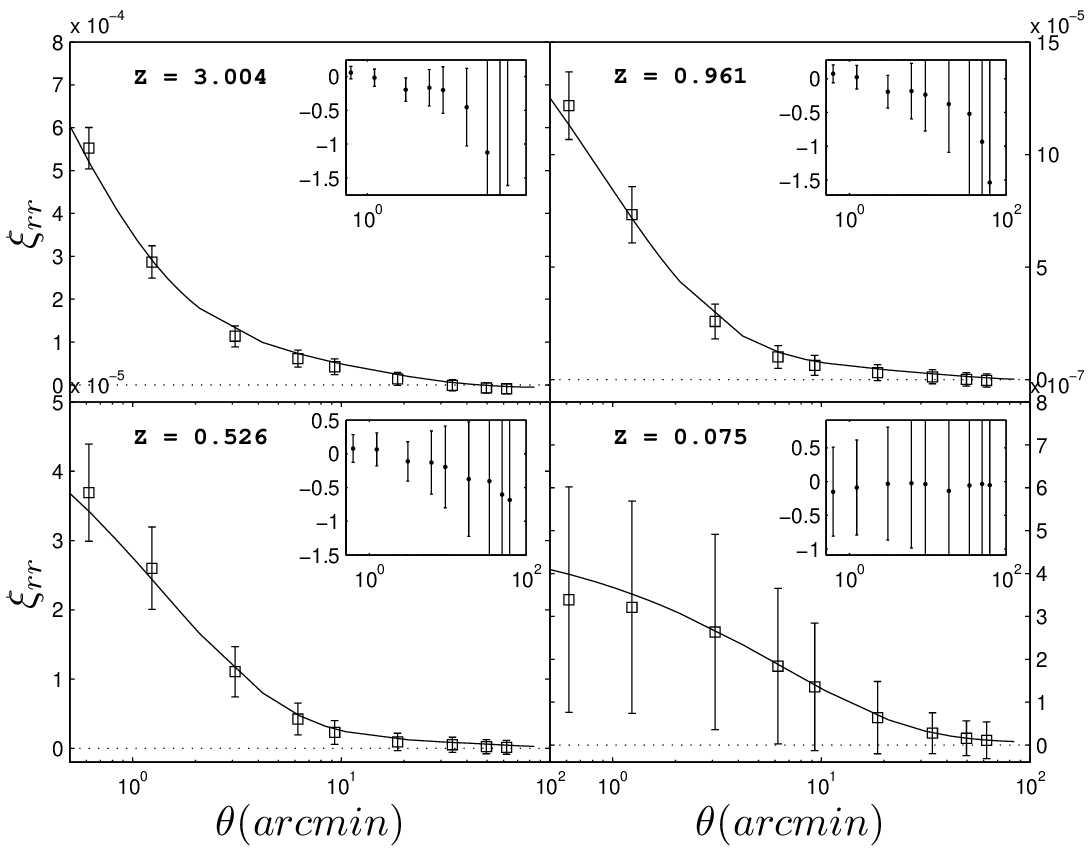}
\end{center}
\caption{$\xi_{rr}$ component. The fractional errors, shown in the insets, 
are larger than for the $tt$ component by about a factor of two, but still consistent at $1\sigma$. 
This is encouraging in the sense that most estimators are based on the latter quantity.
 \label{fig:xirr}}
\end{figure*}

We show in Fig. \ref{fig:xitt} and \ref{fig:xirr} the 2-point correlation functions $\xi_{tt}$ and $\xi_{rr}$ respectively,
as a function of the separation angle and at 4 different redshifts.
The error bars are the $1\sigma$ standard deviation on the sampling, as estimated from our 185 lines-of-sight. 
The agreements between the simulations and  the theoretical predictions are well within the error bars down to $0.6$ arcmin,
which allows us to conclude that the signals are well resolved at least in that range.

\begin{figure*}
\begin{center}
\epsfig{file= 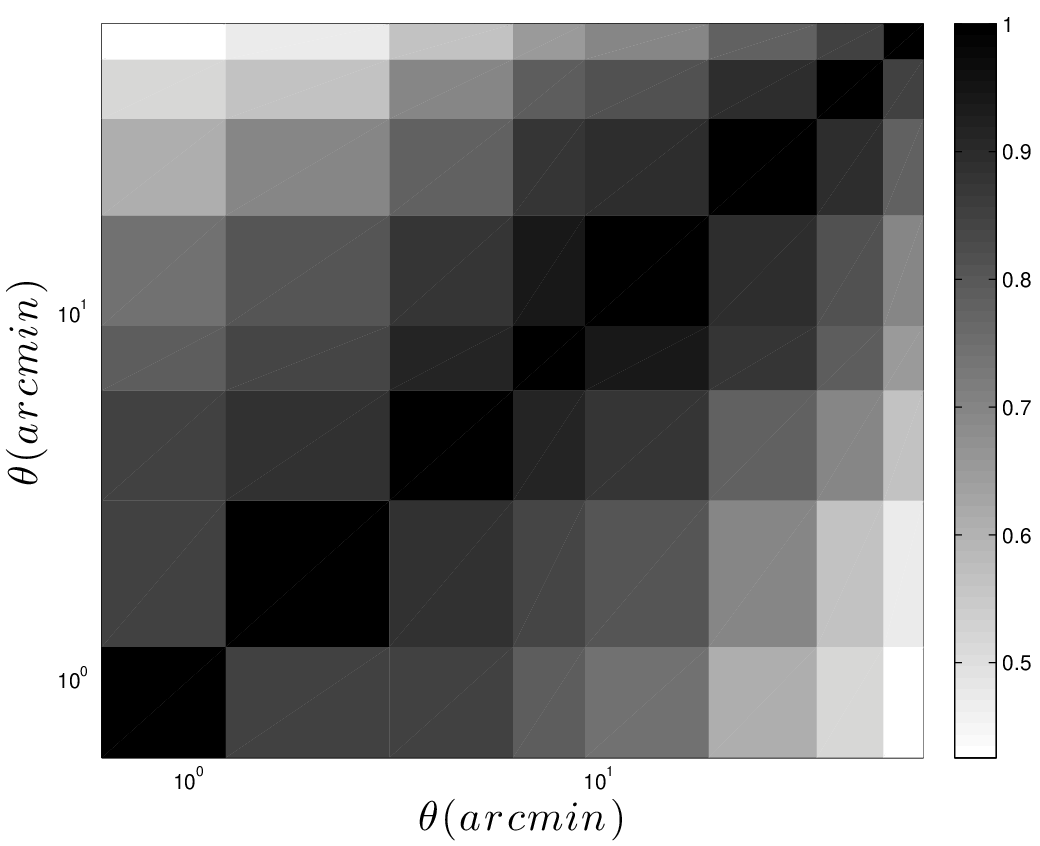,width=0.39\textwidth,height=0.39\textwidth}
\epsfig{file= 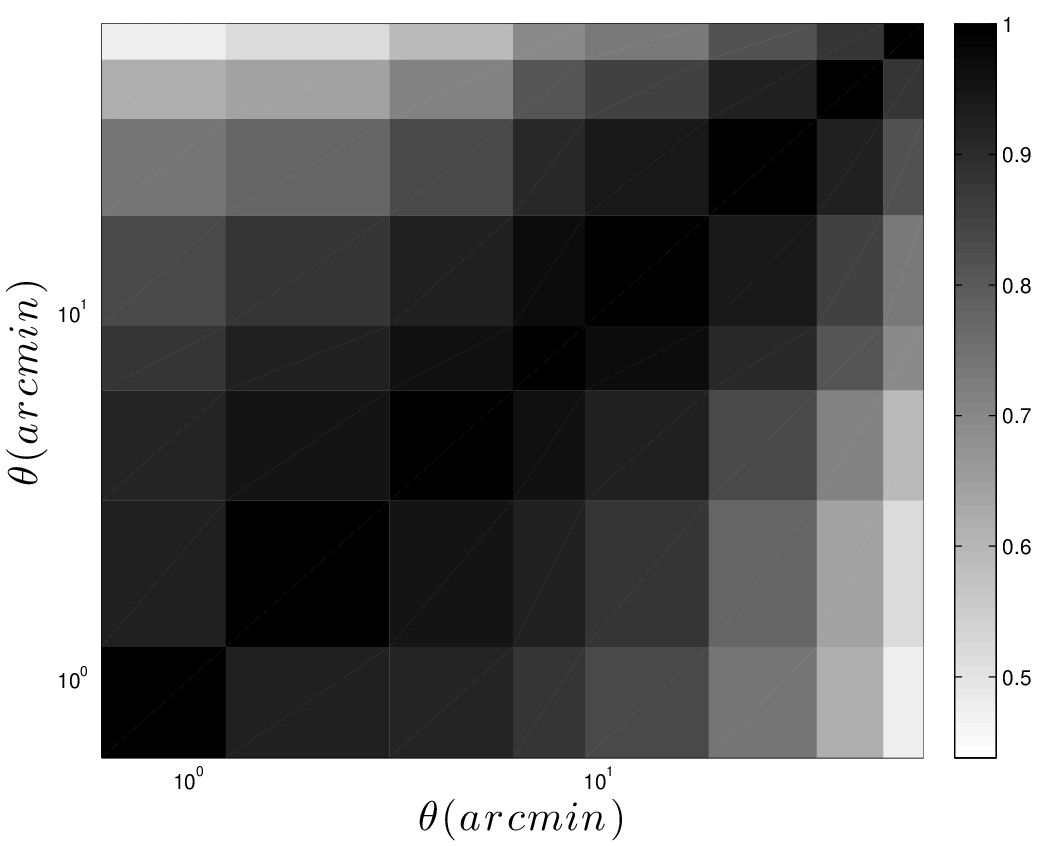,width=0.39\textwidth,height=0.39\textwidth}
\caption{Cross-correlation coefficient matrix of the $tt$ two-point function, with the source plane at  $z \sim 3.0$ (left)  and $z \sim1.0$ (right). These exhibit the strong correlation 
that exists between different angular bins. \label{fig:rho_tt_highZ}}
\end{center}
\end{figure*}

We next show in Fig. \ref{fig:rho_tt_highZ}  
the cross-correlation coefficient matrix related to the $tt$ measurement, 
for source redshifts at 3.0 and 1.0. 
These show that the error bars across different angles are at least 
 50 per cent correlated for the highest redshift, and up to 80 per cent for lower redshift sources. 
 This correlation becomes even stronger as the two angles become similar in size.
 Any robust results based on the uncertainty about $\xi_{tt}$ must therefore incorporate 
 these off-diagonal components. Typically, calculations that combine these measurements
in a noise-weighted scheme will need to invert the full covariance matrix, and the non-Gaussian
error bars thus obtained will generally be smaller than a naive Gaussian treatment.
The $rr$ matrices are qualitatively similar, to Fig.  \ref{fig:rho_tt_highZ}, 
hence we do not show them here\footnote{The covariance matrices of many weak lensing estimators presented in this paper are visually similar, hence we do not present them all.
They are nevertheless made available upon request.}. 
 
\subsection{Convergence}

\begin{figure*}
\begin{center}
\includegraphics[scale=0.70]{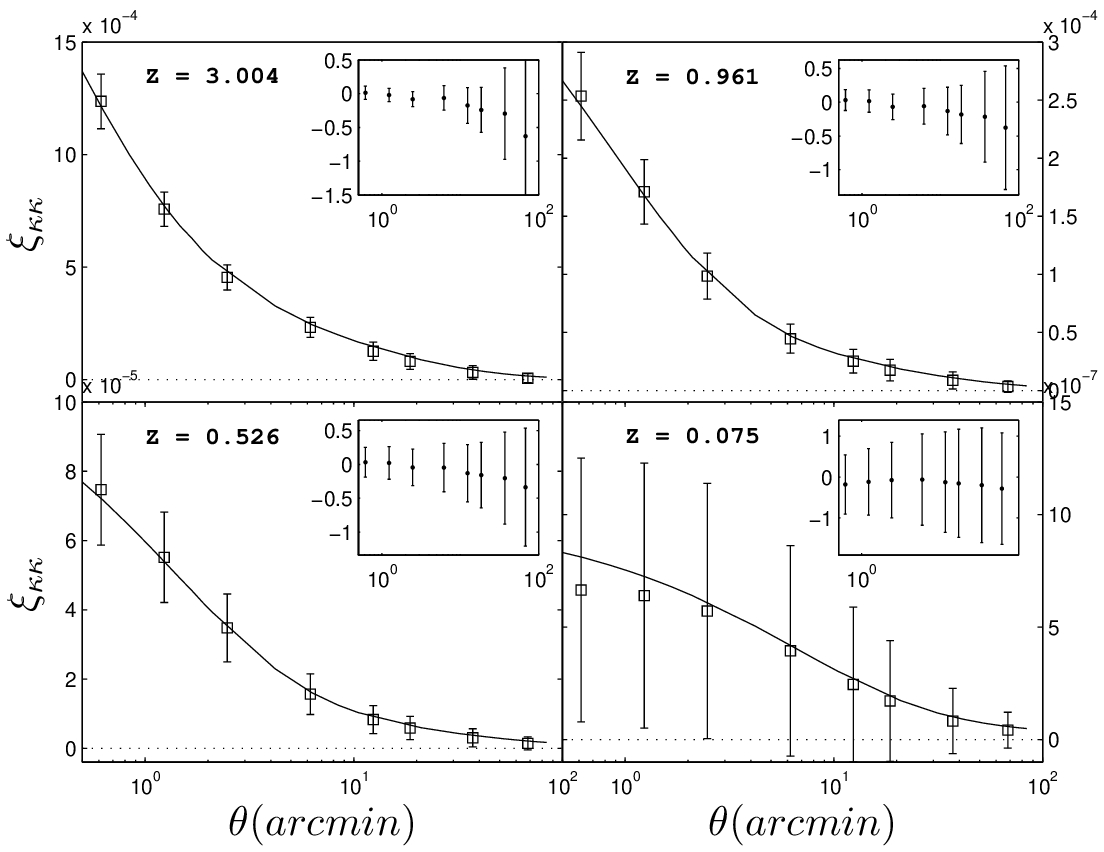}
\end{center}
\caption{Convergence auto-correlation function $\xi_{\kappa \kappa}$, at four different redshifts,
constructed from Poisson samplings the simulated $\kappa$-maps.
Even though the means differ by about 20 per cent, as shown in the insets, 
the agreement with the predictions is well within $1\sigma$. 
\label{fig:kappakappa}}
\end{figure*}

Convergence -- or magnification -- has been successfully detected in recent data analyses  \citep{2005ApJ...633..589S, 2009A&A...507..683H, 2011ApJ...733L..30H}. 
Although more challenging to measure, it serves as an important cross-check of the shear results, plus it is theoretically cleaner:
no (non-local) Fourier transforms are needed in the reconstruction of the underlying dark matter density field.
Following the procedure developed in the last section, we calculate the $\kappa$ auto-correlation function from Poisson sampling the simulated convergence maps. 
In Fig. \ref{fig:kappakappa}, we present our results and find that the agreement with
the non-linear theoretical predictions extends deep under the arc minute at all redshifts.
The cross-correlation coefficient matrices corresponding to these measurements
are very similar to the $\gamma_{tt}$ matrices, hence we do not present the matrices here.

\section{Weak Lensing with Window-Integrated Correlation Functions }
\label{sec:estimators2}

As mentioned earlier, the 2-point correlation  functions are not always the best way to measure the cosmic shear or convergence. Window-integrated correlation functions, such as the mass aperture variance, give a second  handle on many cosmological parameters \citep{Schneider98}, and are also used in galaxy-galaxy lensing and cluster lensing.
The method consists in measuring the mean, the variance or even higher-order moments of a given lensing field, which was beforehand convolved with a filter of variable smoothening angle $\theta$, generally either a `top hat' or a compensated filter. The integrated results are then computed as a function of $\theta$.

When extracting such estimators from the shear signal, the choice of filter matters. One of the advantage of the top hat filter is that it probes scales as large as the field of view,
whereas compensated filters are limited by a damping tail in the window shape that somehow wastes the boundaries of the patch. Another advantage of the top hat filter is that it yields a signal-to-noise ratio that is optimal for skewness measurements \citep{Vafaei10}. On the other hand, a compensated mass aperture filter is more sensitive to small scales, hence it does a better jobs at recovering the signal from surveys with limited sky coverage.
In addition, it is measured directly from the tangential shear field \citep{1993ApJ...404..441K, 1996MNRAS.283..837S}, hence is not affected by the systematic and statistical uncertainties involved in the reconstruction of the convergence field.

\begin{figure*}
\begin{center}
\includegraphics[scale=0.70]{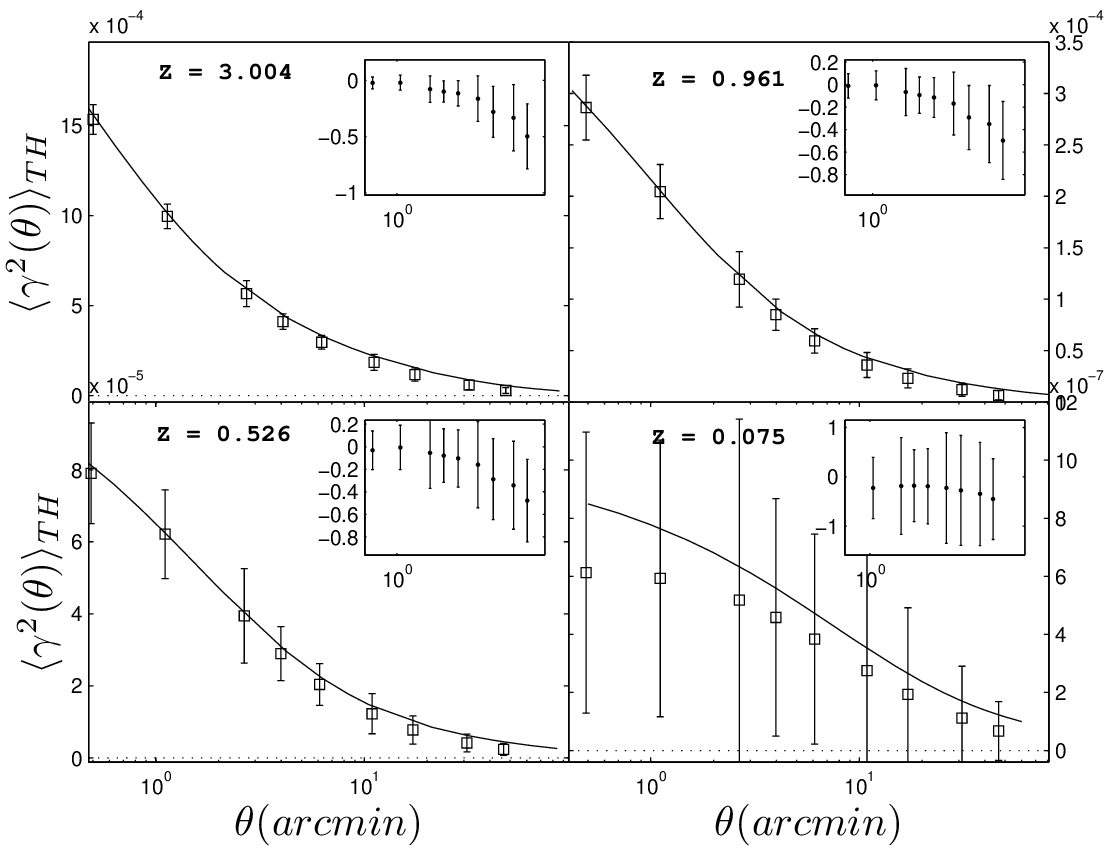}
\end{center}
\caption{Top hat variance,  $\langle |\bar{\gamma}|^2 \rangle$,  measured from shear maps, at 4 different redshifts. There is a small bias in the mean of our measurements, especially at low redshift, however the predictions are still well within the error bars. \label{fig:intTH}}
\end{figure*}

\begin{figure*}
\begin{center}
\epsfig{file= 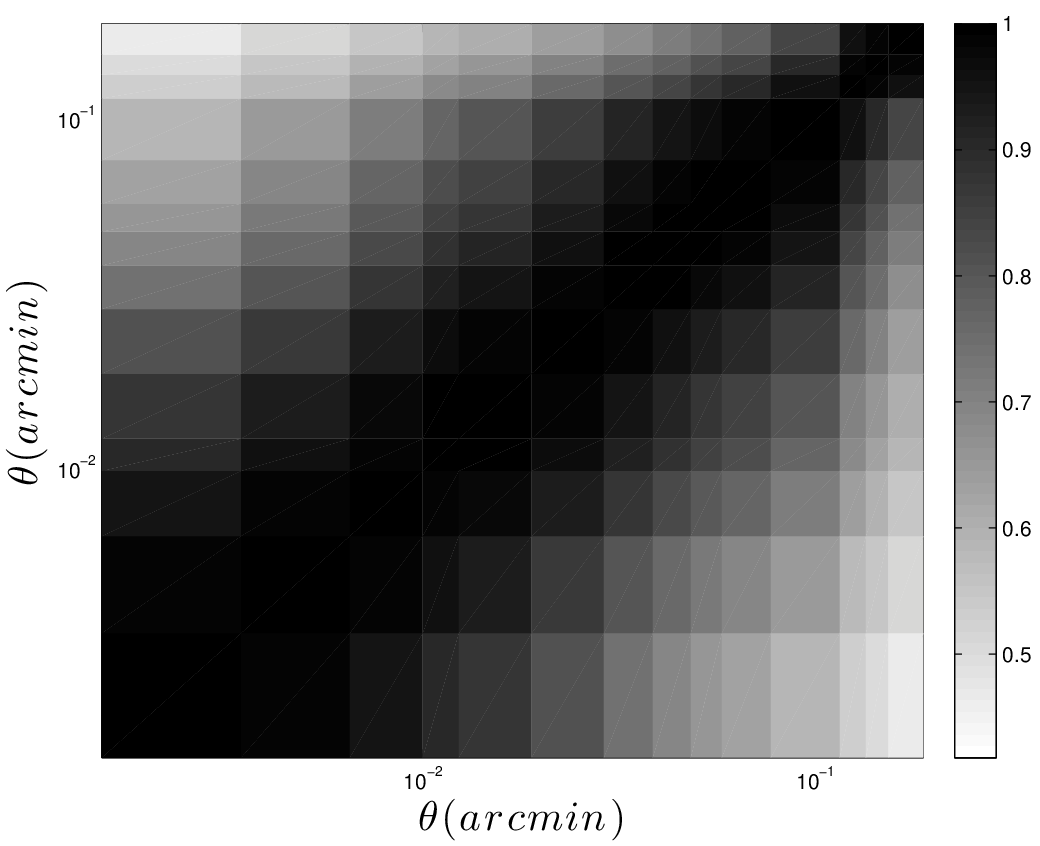,width=0.39\textwidth,height=0.39\textwidth}
\epsfig{file= 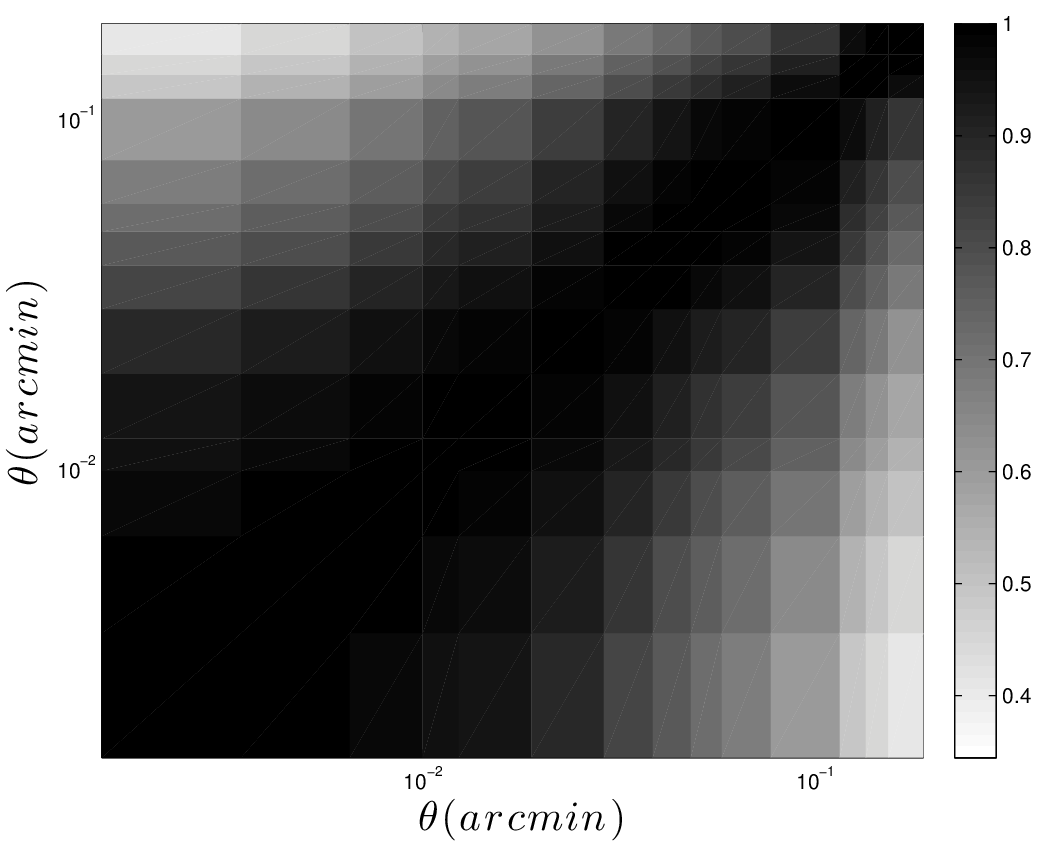,width=0.39\textwidth,height=0.39\textwidth}
\caption{Cross-correlation coefficient matrix of the top hat variance, with the source plane fixed at $z \sim  3.0$ (left) and  $z \sim1.0$ (right). \label{fig:rho_TH_highZ}}
\end{center}
\end{figure*}

The top hat filter  is a circular aperture of radius $\theta$, outside of which the signal is cut to zero. 
The filtering process then measures the total shear in a filtered region of a map,  $\bar{\gamma}$, for a given opening angle $\theta$. This convolution is performed with Fourier transforms, and each of the maps are zero-padded in order to reduce the effect of boundaries. 
We repeat such measurement over all maps  and compute 
the variance $\langle |\bar{\gamma}|^2\rangle_{TH}$, which is related to the convergence power spectrum  $C_\ell$  as:
\begin{eqnarray}
\langle |\bar{\gamma}^2(\theta)| \rangle_{TH}=\frac{1}{2 \pi}\int d\ell \ell C_\ell W_{TH}(\ell \theta)
\label{eq:tophatvar}
\end{eqnarray}
with $W_{TH}(\ell \theta)=\frac{4J_1^2(\ell \theta)}{(\ell \theta)^2}$ \citep{Kaiser92}.
We compare our measurements  with the non-linear predictions in Fig. \ref{fig:intTH},
as a function of $\theta$,  and find a good agreement at all redshift. 
There is a  small bias in the mean, which is nevertheless consistent within $1\sigma$. 
The cross-correlation matrices associated with two of these measurements 
are presented in Fig. \ref{fig:rho_TH_highZ} and exhibit the strong correlation that exists
between most angular bins.

\begin{figure*}
\begin{center}
\includegraphics[scale=0.70]{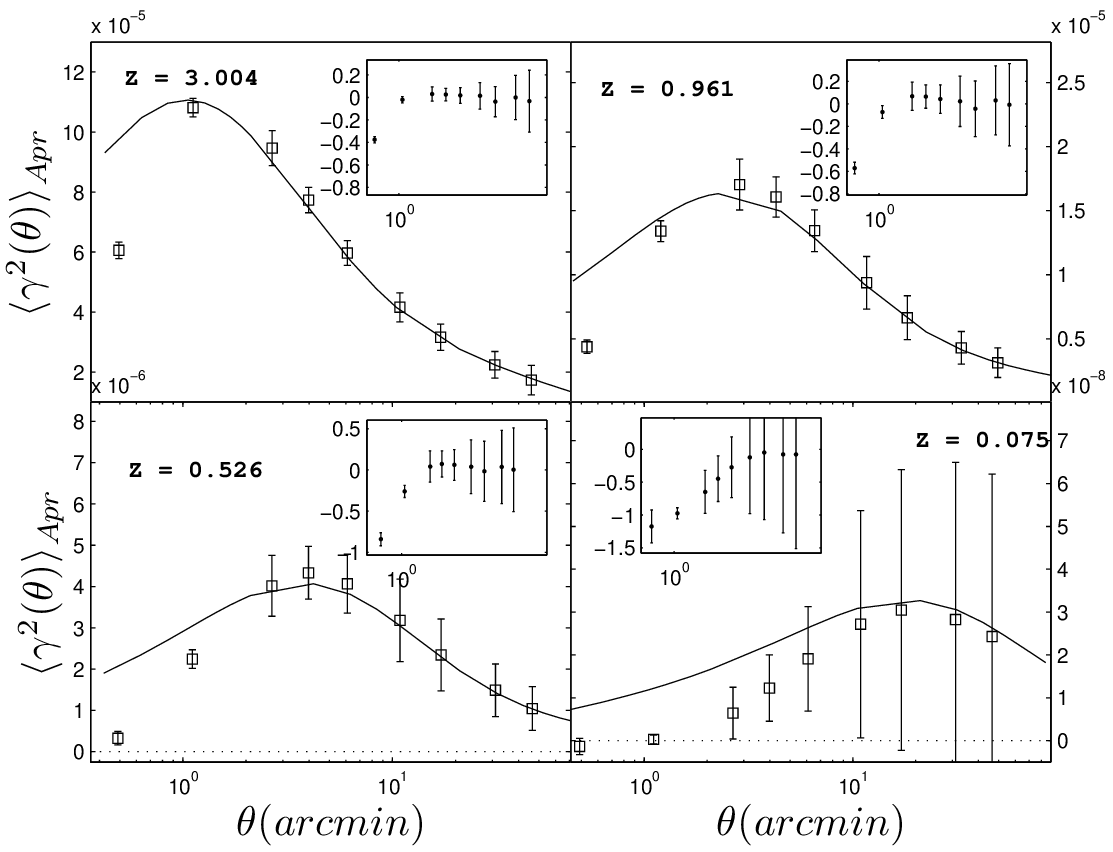}
\end{center}
\caption{Aperture mass variance $\langle M_{ap}^2 \rangle$ measured from tangential shear maps. 
The apparent discrepancy between simulations and theoretical predictions at low redshift is caused by 
resolution limits, where the smallest angles actually probe scales that are approaching the grid size.
The intrinsic pixel size of $0.21$ arcmin correspond to angles of about $1.0$ arcmin with
this compensated filter. 
\label{fig:intApr}}
\end{figure*}

\begin{figure*}
\begin{center}
\epsfig{file= 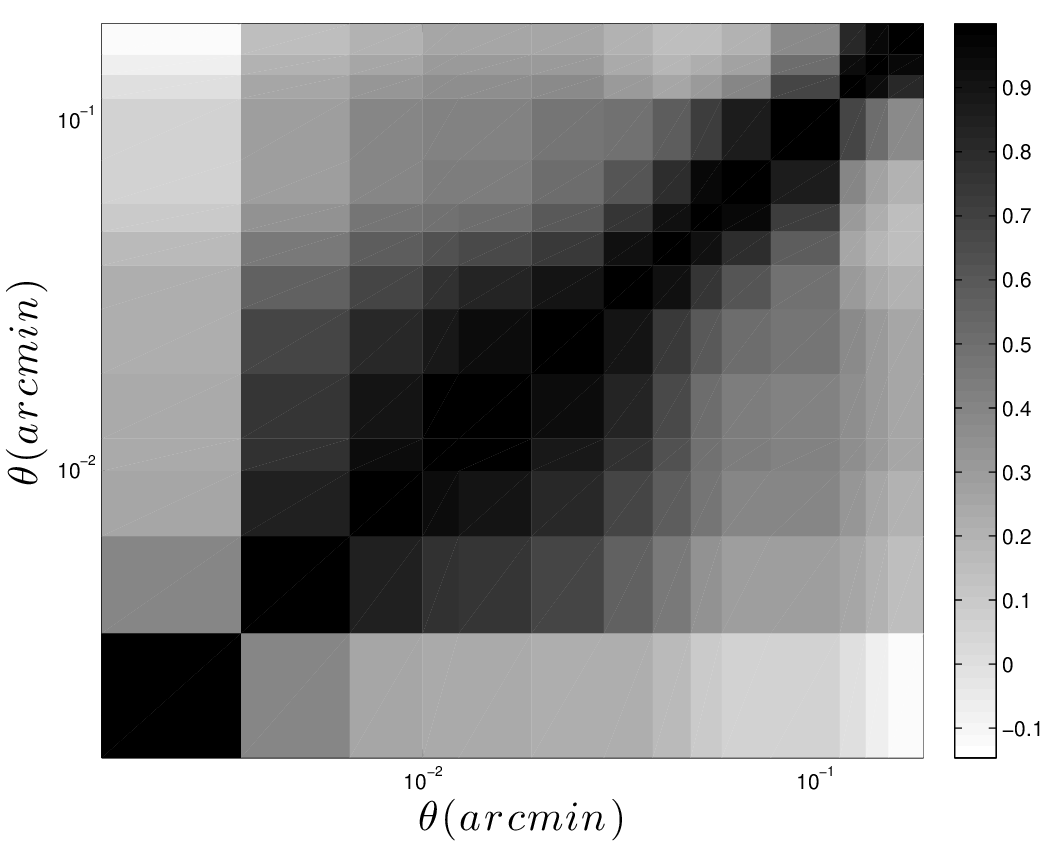,width=0.39\textwidth,height=0.39\textwidth}
\epsfig{file= 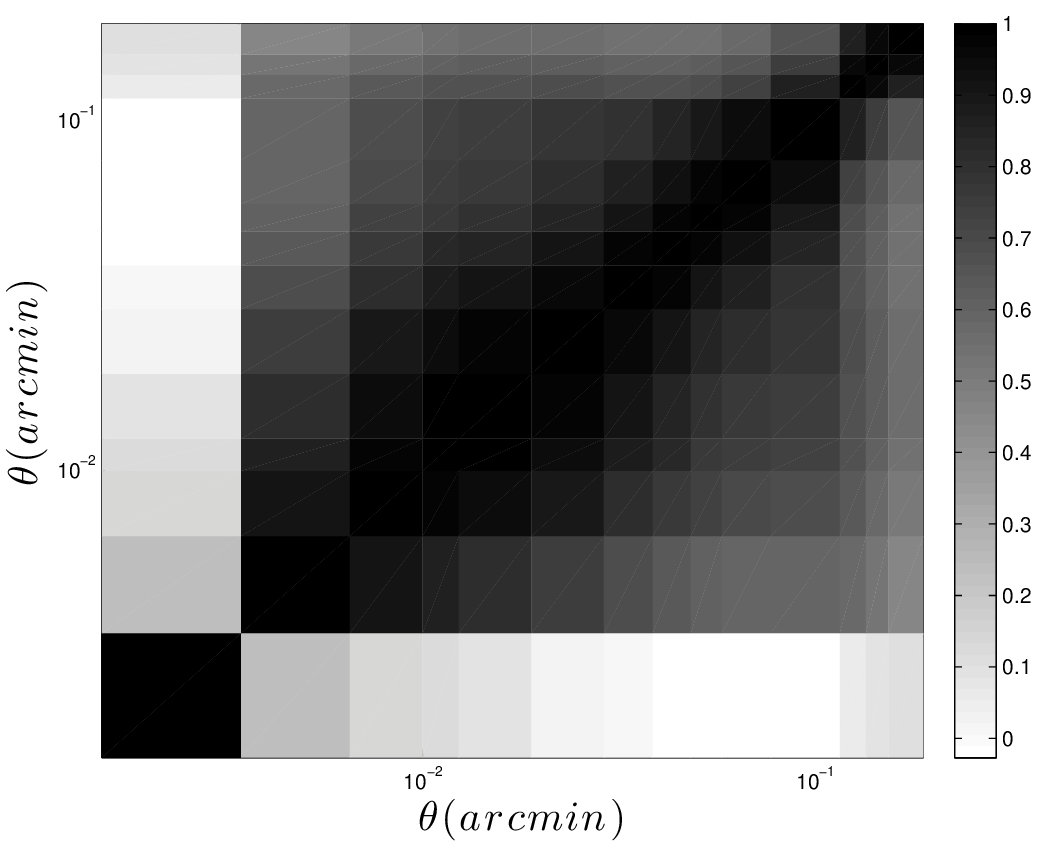,width=0.39\textwidth,height=0.39\textwidth}
\caption{Cross-correlation coefficient matrix of the  aperture mass variance, measured from the tangential shear, with the source plane fixed at $z \sim  3.0$ (left) and  $z \sim1.0$ (right). The correlation seems to vanish in the first bin, which is approaching
the simulation resolution limit.  \label{fig:rho_Apr_highZ}}
\end{center}
\end{figure*}

We next consider a compensated aperture filter, which is constructed from the local tangential shear mock catalogues. In this process, one of the galaxy in the pair is replaced by the  centre of the filter. 
The aperture mass $M_{ap}$ is then given by \citep{Schneider98}:
\begin{eqnarray}
M_{ap}(\theta)=\int d^2 \vartheta Q_{\theta}(\vartheta)\gamma_t(\vartheta)
\label{eq:Map}
\end{eqnarray}
where $Q$ is a weight function with support $|\vartheta| \in [0,\theta]$ and which takes the shape:
\begin{eqnarray}
Q_{\theta}(\vartheta)=\frac{6}{\pi \theta^2}\left( \frac{\vartheta}{\theta^2}\right) \left( 1-\frac{\vartheta^2}{\theta^2}\right)
\label{eq:Q}
\end{eqnarray}
We then calculate the variance $\langle M_{ap}^2 \rangle$ across the map, for all available angles, 
which is also related to the convergence power spectrum:
\begin{eqnarray}
\langle M_{ap}^2 (\theta)\rangle=\frac{1}{2 \pi}\int_0^\infty d\ell \ell C_{\ell} W_{ap}(\ell \theta)
\label{eq:mapvar}
\end{eqnarray}
with $W_{ap}(\ell \theta)=\frac{276 J_4^2(\ell \theta)}{(\ell \theta)^4}$.
We present in  Fig. \ref{fig:intApr}  our measurements of $\langle M_{ap}^2 \rangle $ from the simulations, 
as a function of smoothing scale $\theta$. Over-plotted are the theoretical predictions obtained from [Eq.  \ref{eq:mapvar}].
We observe that for  $z>1.0$, the agreement extends down to the arc minute,
whereas lower redshifts suffer from a lack of variance at angles of a few arc minutes.
This is caused by limitations in the resolution due to strong zooming from the simulation grid
on to the pixel map. 
We recall that with compensated filters, an opening angle $\theta$ really probes scales
at an angle $\sim \theta/5$, which approach the simulation resolution at very low redshifts.
This drop is also expected from the top hat variance, but appears at much smaller smoothing angles.
The cross-correlation coefficient matrices are presented in Fig. \ref{fig:rho_Apr_highZ},
and show that most measurements are close to 60 per cent correlated. The smallest angles
probe scales that approach the pixel resolution, hence there is very little cross-correlation.

\section{Weak  Lensing with Windowed Statistics on Convergence Maps}
\label{sec:estimators3}

Window statistics performed directly on the convergence fields serve as an important test of the accuracy and precision of the simulations,
since the calculations here can be done directly on the grid, i.e. without the Poisson sampling. 
We smooth the $\kappa$-maps with filters identical to those used in the last section and
calculate the top hat  variance $\langle \bar{\kappa}^{2}(\theta)\rangle_{TH}$ and mass aperture variance $\langle M_{ap}^{2}(\theta)\rangle$ as a function of the filter  opening angle.
In the latter case, the choice of compensated filter ( i.e. the equivalent of $Q_\theta(\vartheta)$ in [Eq. \ref{eq:Q}] )
is given by $U_\theta(\vartheta)$ \citep{Schneider98}, where:
\begin{eqnarray}
U_{\theta}(\vartheta)=\frac{9}{\pi \theta^2} \left( 1-\frac{\vartheta}{\theta^2}\right)\left(\frac{1}{3}-\frac{\vartheta^2}{\theta^2}\right)
\label{eq:apr_ft}
\end{eqnarray}
The $M_{ap}$ estimator  is now obtained from the convergence maps as:
\begin{eqnarray}
M_{ap}(\theta)=\int d^2 \vartheta U_{\theta}(\vartheta)\kappa(\vartheta)
\label{eq:Map_kappa}
\end{eqnarray}
This measurement is complimentary to the shear approach (Eq. \ref{eq:Map}), 
and will yield identical results if the systematics are well understood.

\begin{figure*}
\begin{center}
\includegraphics[scale=0.70]{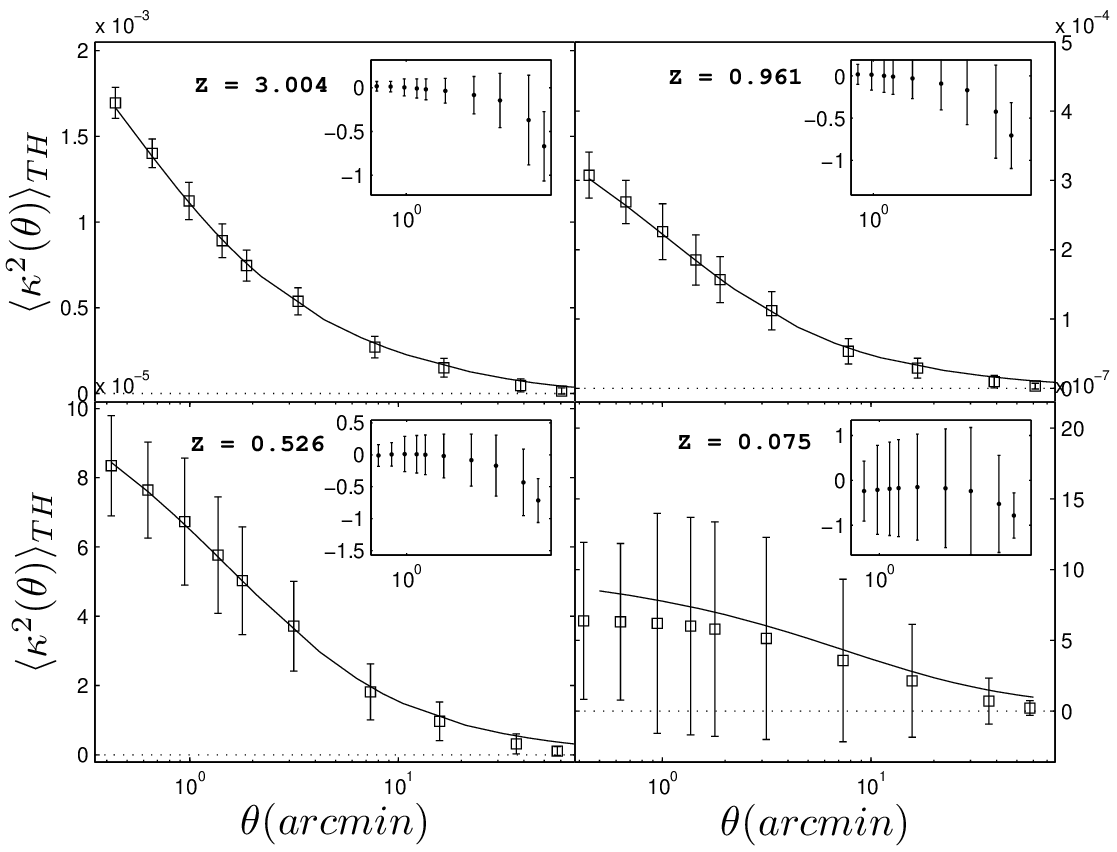}
\end{center}
\caption{ $\langle |\bar{\gamma}|^2 \rangle_{TH}$  measured directly from the convergence maps. 
We see that the agreement with the theoretical predictions is  excellent at all redshifts. 
In absence of large systematic uncertainties, as prevails in simulated environments, this figure 
is similar to Fig. \ref{fig:intTH}.  \label{fig:windowTH2}}
\end{figure*}

\begin{figure*}
\begin{center}
\includegraphics[scale=0.70]{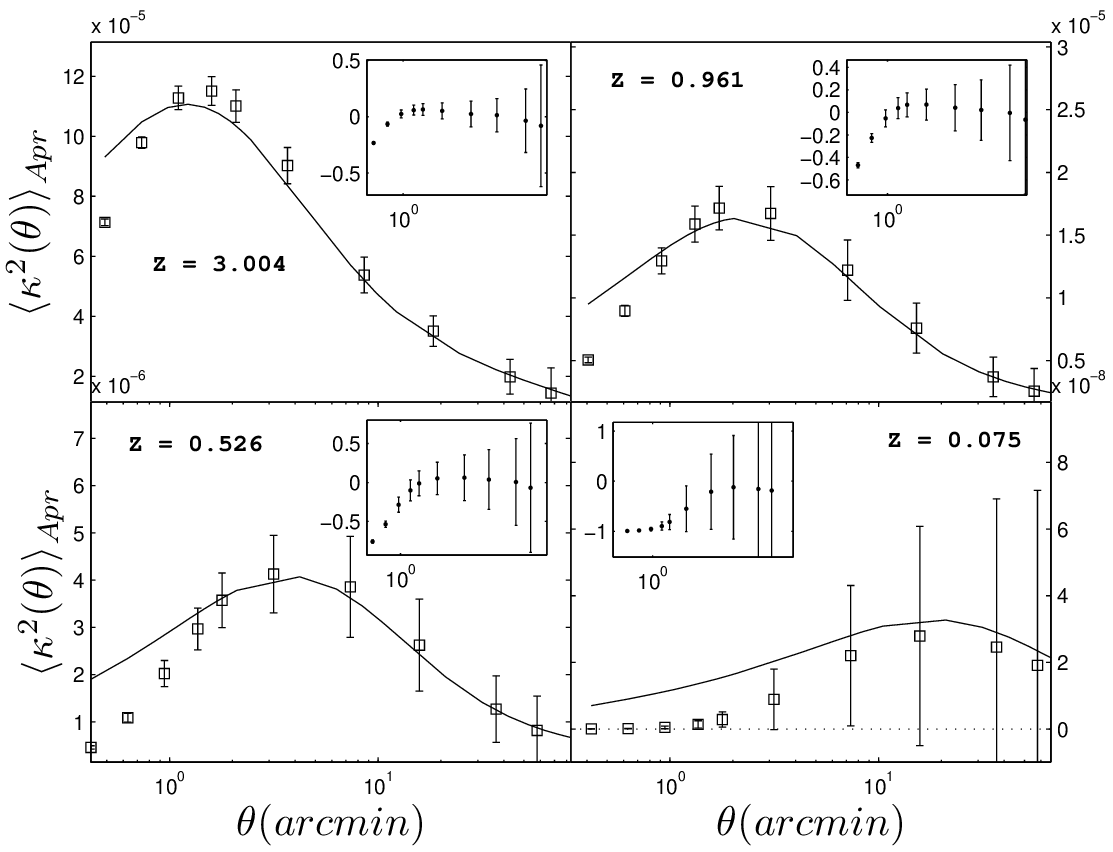}
\end{center}
\caption{Aperture mass variance $\langle M_{ap}^2 \rangle$, measured directly from the convergence maps.
We recall that the effect of finite pixel size is felt at larger angular scales -- up to about one arc minute -- with this estimator. This figure is equivalent to Fig. \ref{fig:intApr}. 
\label{fig:windowApr2}}
\end{figure*}

We present in Fig. \ref{fig:windowTH2} and \ref{fig:windowApr2}  
our measurements of the top hat and mass aperture variance respectively.
We observe that the signals are almost identical with  the corresponding shear estimators
(Fig. \ref{fig:intTH} and \ref{fig:intApr}).
The agreement with the predictions is good at all redshifts for the top hat variance, with only a slight bias at the lowest redshifts, 
whereas the mass aperture variance shows a lack of signal  at angles of a few arc minutes for low redshifts, consistent with the shear results.

\begin{figure*}
\begin{center}
\includegraphics[scale=0.70]{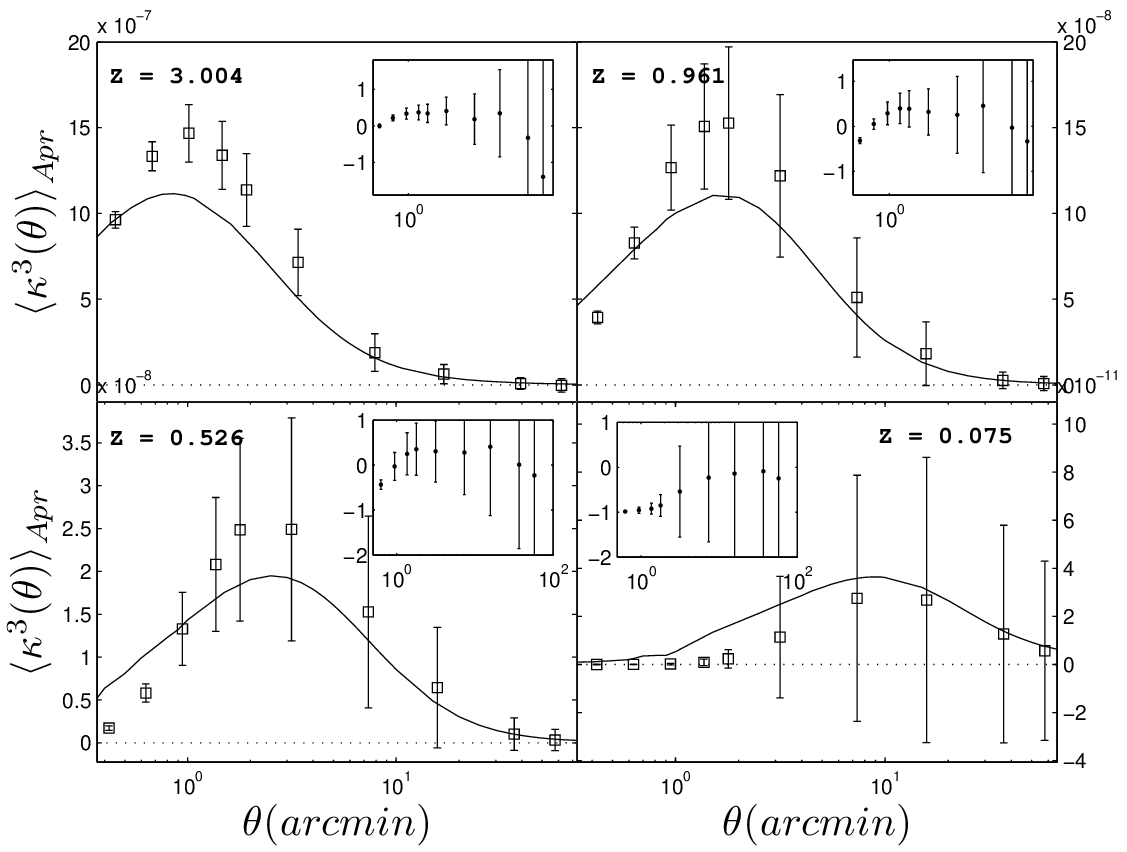}
\end{center}
\caption{$\langle M_{ap}^3 \rangle$ measured directly from the convergence maps. 
We recall that the effect of finite pixel size is felt to larger angular scales -- up to about one arc minute in lower redshift lenses -- with this estimator,
which explains the damped tail at small angles in the $z=0.075$ plot.
 \label{fig:windowApr3}}
\end{figure*}

\begin{figure*}
\begin{center}
\includegraphics[scale=0.70]{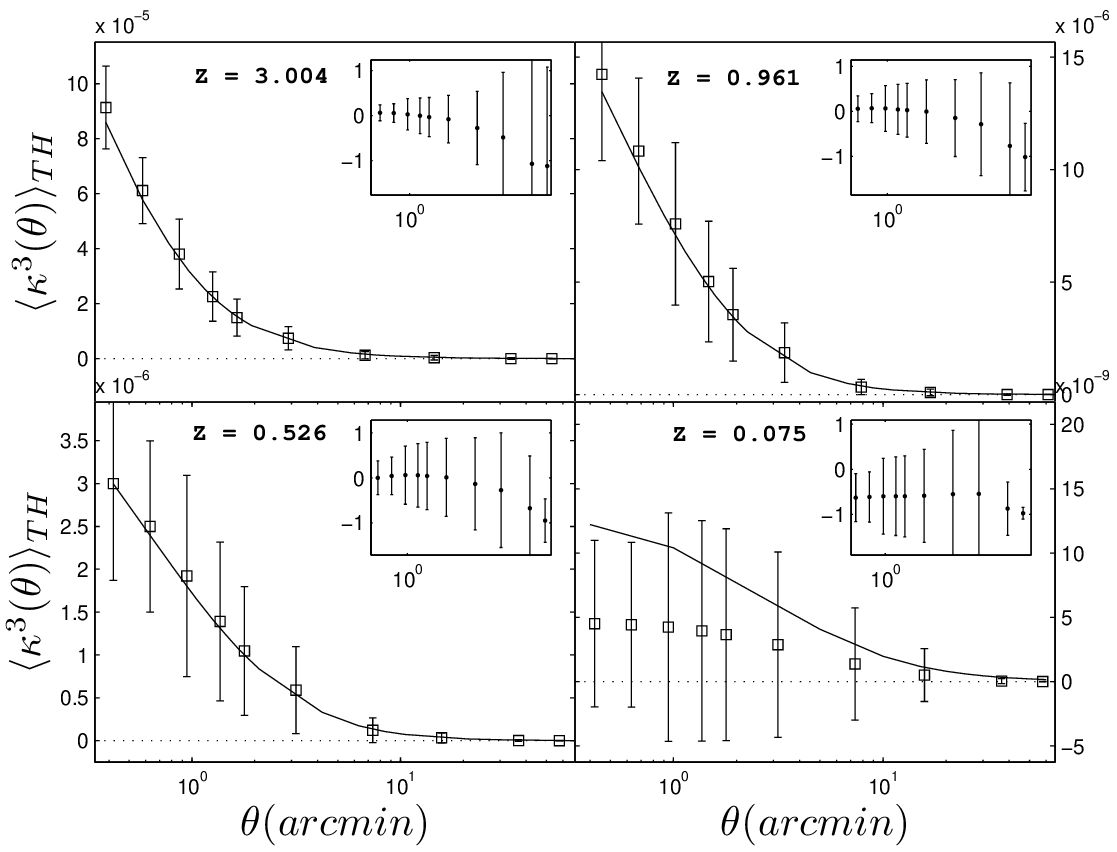}
\end{center}
\caption{$\langle |\bar{\gamma}|^3 \rangle_{TH}$ measured directly from the convergence maps.\label{fig:windowTH3}}
\end{figure*}

We finally show the three-point function $\langle M_{ap}^{3}(\theta)\rangle$  and $\langle \bar{\kappa}^{3}(\theta) \rangle_{TH}$ in Fig. \ref{fig:windowApr3} 
and \ref{fig:windowTH3}, both measured directly on the convergence maps. 
We recall that these measurements are essential to break the degeneracy between $\sigma_8$ and $\Omega_m$.
We observe a good agreement between the predictions and the top hat measurements,
whereas the simulations tend to overestimate the mass aperture predictions by $1\sigma$ at low angles.
This comes again from the fact that the aperture filter is sensitive to about one fifth of the total opening angle probed.
Hence the discrepancy observed at $\theta \sim 2$ arcmin is mainly probing scales of $0.4$ arcmin, which is of the order of the pixel size.

\section{Conclusion}
\label{sec:conclusion}

This paper has two principal objectives: 1) measure the non-Gaussian covariance matrix
on the principal weak lensing estimators with sub-arc minute precision , and 2) set the stage for systematic studies of secondary effects, and especially how 
their combination impacts the lensing signal.
We have generated a set of $185$ high resolution N-body simulations, the {\small TCS} simulation suite, from which we constructed
past light cones with a ray tracing algorithm. The weak lensing signal is accurately resolved 
from a few degrees down to a fraction of an arc minute. Thanks to the large statistics, we have measured non-Gaussian error bars on
a variety of weak lensing estimators, including 2-point correlation functions on shear and convergence maps, 
and window-integrated estimators such as the mass aperture.
In each case, we compared our results with non-linear theoretical predictions at a few redshifts and obtained a good agreement,
 which testifies the quality of the simulations.
 
 In addition, we measured  the covariance matrices for each of these estimators,
and we show that the error bars between most angular measurements are at least $50$  per cent
correlated, with regions up to $90$ per cent correlated, especially when the two angles become closer. 
These non-Gaussian, correlated, error bars are essential for a correct estimate of many derived quantities -- 
including  cosmological parameters like $\sigma_8$, $\Omega_m$ or $w$, 
which so far relied either on Gaussian assumptions, or on numerical estimates that were not resolving the complete dynamical range. 
With the next generation of lensing survey, however, these non-Gaussian error bars, which intrinsically deviate significantly from Gaussian prescriptions,
are expected to be resolved, therefore techniques such as those presented here
will be required for robust estimates.

We also generated a series of halo mock catalogues that are coupled to the gravitational lenses 
for future independent studies of secondary signals and alternate tests of weak lensing estimators.
Within the CFHTLenS collaboration, these catalogues will be part of the {\small CLONE} project\footnote{The {\small CLONE} and the weak lensing maps are available for download at  {\tt vn90.phas.ubc.ca/jharno/CFHT\_Mock\_Public/}. The halo catalogues are available upon request.}. 
Our near term goal is to include effects such as intrinsic alignment, source clustering, etc. in a single galaxy population algorithm
and quantify their combined contribution. In addition, we plan to quantify the impact of post-Born calculations on the non-Gaussian uncertainty and   
on the contamination by secondary signals. Understanding the impact of all these effects is essential, for such systematic bias
are likely to contribute significantly to future surveys such as KiDS and Euclid.

Aspects not included in our simulation settings are baryon feedback effects \citep{2011MNRAS.417.2020S} and dependence of the covariance matrices on the cosmological parameters \citep{2009A&A...502..721E}. While the latter can be simply addressed by running additional simulations,  
the former requires hydrodynamical simulations that implement simultaneously matter clustering at large angular scales and a proper modelling of feedback effects.

\section*{Acknowledgments}
The authors would like to thank Ue-Li Pen, Dmitri Pogosyan, Catherine Heymans, Fergus Simpson, Elisabetta Semboloni, Hendrick Hildebrandt,
 Martin Kilbinger  and Christopher Bonnett for useful discussions and comments on the manuscripts.
LVW acknowledges  the financial support of NSERC and CIfAR, and JHD is grateful for the FQRNT scholarship 
under which parts of this research was conducted.
N-body computations were performed on the TCS supercomputer at the SciNet HPC Consortium. SciNet is funded by: the Canada Foundation for Innovation under the auspices of Compute Canada; the Government of Ontario; Ontario Research Fund - Research Excellence; and the University of Toronto.

\bibliographystyle{mn2e}
\bibliography{lensing}{}

\bsp

\label{lastpage}

\end{document}